\documentclass[twocolumn,showpacs,preprintnumbers,amsmath,amssymb,aps,prl,10pt]{revtex4-1}
\usepackage{graphicx}
\usepackage{dcolumn} 
\usepackage{bm} 
\usepackage{longtable}
\usepackage{color} 
\usepackage[normalem]{ulem} 
\usepackage{amssymb} 
\usepackage{bbm} 
\usepackage{upgreek} 
\usepackage{lipsum}
\usepackage[utf8]{inputenc}
\graphicspath{{./}}

\begin{document}
\title{Spin-, time- and angle-resolved photoemission spectroscopy on WTe$_2$}
\author{Mauro Fanciulli$^{1,2}$, Jakub Schusser$^{1,2,3}$, Min-I Lee$^{1,2}$, Zakariae El Youbi$^{1,2,4}$, Olivier Heckmann$^{1,2}$, Maria Christine Richter$^{1,2}$, Cephise Cacho$^{4}$, Carlo Spezzani$^{5}$, David Bresteau$^{2}$, Jean-François Hergott$^{2}$, Pascal D'Oliveira$^{2}$, Olivier Tcherbakoff$^{2}$, Thierry Ruchon$^{2}$, Jan Minár$^{3}$, Karol Hricovini$^{1,2}$}
\affiliation{$^{1}$Laboratoire de Physique des Matériaux et Surfaces, Université de Cergy-Pontoise, 95031 Cergy-Pontoise, France
\\
$^{2}$Laboratoire d'Interactions, DYnamiques et Lasers, CEA, CNRS, Université Paris-Saclay, CEA Saclay, 91191 Gif-sur-Yvette, France
\\
$^{3}$New Technologies-Research Center, University of West Bohemia, 30614 Pilsen, Czech Republic
\\
$^{4}$Diamond Light Source, Harwell Campus, OX110DE Didcot, United Kingdom
\\
$^{5}$Elettra-Sincrotrone Trieste, 34149 Basovizza, Italy}
\date{\today}
\begin{abstract}
We combined a spin-resolved photoemission spectrometer with a high-harmonic generation (HHG) laser source in order to perform spin-, time- and angle-resolved photoemission spectroscopy (STARPES) experiments on the transition metal dichalcogenide bulk WTe$_2$, a possible Weyl type-II semimetal.
Measurements at different femtosecond pump-probe delays and comparison with spin-resolved one-step photoemission calculations provide insight into the spin polarization of electrons above the Fermi level in the region where Weyl points of WTe$_2$ are expected.
We observe a spin accumulation above the Weyl points region, that is consistent with a spin-selective bottleneck effect due to the presence of spin polarized cone-like electronic structure.
Our results support the feasibility of STARPES with HHG, which despite being experimentally challenging provides a unique way to study spin dynamics in photoemission.
\end{abstract}
\maketitle
\begin{figure} [ht!]
	\centering
		\includegraphics[width=0.48\textwidth]{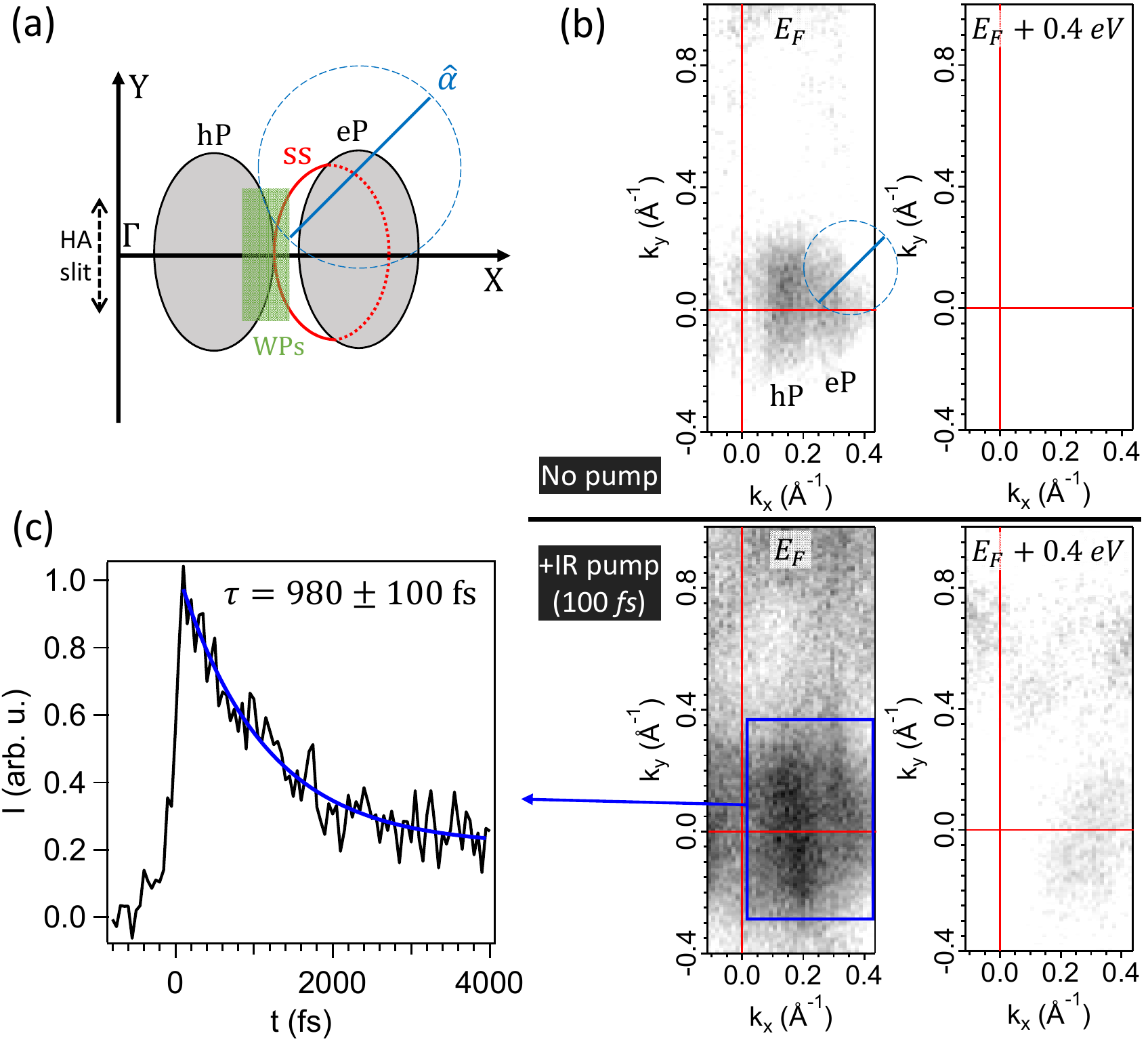}
	\caption{(a) Schematics of the Fermi surface of WTe$_2$, with a bulk hole pocket (hP) and electron pocket (eP) in black, and a surface state (SS) in red. The region where WPs are expected is shown in green. The hemispherical analyzer (HA) slit direction, the spin quantization axis $\hat{\alpha}$ in the $xy$ plane and the angular resolution (blue circle) of spin resolved data of Fig.~\ref{fig:fig2} are shown. (b) CEMs measured with $h\nu=35.65$~eV for two energies, $E_F$ (left) and $400$~meV above (right), in the situation without pump (top) and $100$~fs after a pump pulse (bottom). (c) Integrated intensity in the angular range shown by the blue rectangle in (b) and an energy range between $E_F$ and $500$~meV above it.}
	\label{fig:fig1}
\end{figure}

WTe$_2$ is a well-studied semimetal belonging to the class of transition metal dichalcogenides.
It has attracted a lot of interest since it presents a non-saturating linear anomalous magnetoresistance \cite{Ali:2014}, pressure-induced superconductivity \cite{Pan:2015} and it is the first proposed topological type-II Weyl semimetal \cite{Soluyanov:2015}.
The existence of semimetallic materials with symmetry-protected linear crossings of bands was pioneered by Abrikosov \cite{Abrikosov:1971}, but the recent reinterpretation in terms of topology provides a new point of view.
Weyl points (WPs) are topologically protected crossings of electron and hole states, which come in pairs of opposite chirality, and are connected by topological surface states dubbed as Fermi arcs. WPs host low-energy excitations that can be described as Weyl fermions. Since Lorentz invariance is not a requirement for collective quasiparticles in condensed matter, one can have inequivalent velocities for the Weyl fermions along different directions as manifested by strongly tilted cones at the crossing states, which is what distinguishes type-II from type-I Weyl semimetals. 
Similarly to the related material MoTe$_2$ \cite{Tamai:2016, Weber:2018}, the expected number of WPs couples and their precise position in reciprocal space strongly depends on details of the calculations and on small variations of the buckled quasi-2D crystal structure, which makes us rethink the meaning of topological protection.
For WTe$_2$ in particular, its electronic structure is actually close to a topological transition \cite{Russmann:2018, Lv:2017, Zhang:2017} between the topologically trivial and type-II Weyl semimetal phases depending on different parameters such as lattice constant, pressure, temperature, or electron doping. 
Since WPs occur at the crossing of bulk states their projection in the surface Brillouin zone is necessarily hidden in the bulk continuum. Furthermore, the WPs are expected slightly above the Fermi level $E_F$.
Despite these issues, the topological classification of MoTe$_2$ as type-II Weyl semimetal has been established \cite{Tamai:2016}, even though the number of WPs is not clear. 
Instead, for WTe$_2$ the situation remains particularly controversial, since WPs are extremely close to each other in reciprocal space and energy \cite{Soluyanov:2015}.
Fig.~\ref{fig:fig1}~(a) shows a simplified schematics of the projection on the Fermi surface of the occupied part of the bulk band structure of WTe$_2$, where hole pockets (hP) and electron pockets (eP) are shown with black contours. The red line indicates a topologically trivial surface state (SS) that becomes a surface resonance when it overlaps with the eP (dotted line) \cite{Russmann:2018}. The region where one or two couples of WPs extremely close to each other are expected above $E_F$ is shown by a green rectangle. 
For a comment about the spread ambiguity in literature about the nomenclature of Fermi arcs see Ref.~\cite{SOM:}.

The electronic structure of WTe$_2$ has been extensively studied \cite{Wang:2016, Das:2016, Wu:2016, Bruno:2016, DiSante:2017, Das:2019} by angle-resolved photoemission spectroscopy (ARPES), but given its complexity there is no consensus on its topological classification also from the experimental point of view, because small variations in the sample preparation procedure might give different results. 
However the discussion is not crucial, since interesting macroscopic transport properties are definitely present and do not strictly rely on the topological classification, but rather on the peculiarities of the overall electronic structure.
For example, magnon emission in spin-polarized transport was explained with the spin texture of the trivial SS \cite{Kononov:2018}, despite being considered as topological, and similarly an enhanced spin-orbit torque effect was explained with the overall spin texture of the Fermi surface \cite{Li:2018}, while being ascribed to only the Fermi arcs.
Indeed, while mathematically a material can only be either topologically trivial or non-trivial, the phase transition between the two is smooth and continuous for the electronic structure observables \cite{Russmann:2018}.
That is to say, a non-zero but very small energy gap at the WPs will not substantially affect the bulk and surface states shape and spin texture and the overall electron and spin dynamics in the cone region.

ARPES studies on WTe$_2$ have also been extended to the spin and time domains. Spin-resolved ARPES (SARPES) measurements have confirmed the lifting of the spin degeneracy of the occupied states due to the non-centrosymmetric crystal structure \cite{Feng:2016, Das:2016}, and time-resolved ARPES (T-ARPES) measurements have determined electron and hole dynamics with comparable time constant of $1$~ps \cite{Caputo:2018}. The studies present different interpretations to explain the anomalous magnetoresistance.

Since both SARPES and T-ARPES are time-consuming techniques, 
only few attempts have been made to combine them \cite{Scholl:1997, Cinchetti:2006, Weber:2011, Cacho:2015, Barriga:2016, Battiato:2018}, 
and recently using a free electron laser source \cite{Fognini:2014} or high harmonic generation (HHG) laser sources \cite{Plotzing:2016, Eich:2017, Nie:2019}, which additionally provide tunability of the probing photon energy.

In this paper, we report the first experimental results on WTe$_2$ with spin-, time- and angle-resolved photoemission spectroscopy, which we shall call STARPES, obtained with an HHG source. The aim is to explore the spin dynamics in the energy-momentum region of the expected WPs. Our results are supported by \textit{ab initio} photoemission calculations.\\

The experiment was performed at Attolab FAB10, a recently commissioned HHG beamline based on a Ti:Sa laser system ($1.55$~eV) at $10$~kHz repetition rate \cite{Golinelli:2017,Golinelli:2019}. 
The beam is split in two parts: one is used as a pump beam, the other one drives a HHG source. 
A time-preserving monochromator \cite{Frassetto:2011} selects about $250$~meV from the HHG spectrum, obtaining a UV probe beam with pulse duration of about $30$~fs \cite{Grazioli:2014}. 

The bulk single crystal is commercially available (HQ Graphene) and was cleaved \textit{in situ} by scotch tape at a base pressure of $10^{-10}$~mbar. The measurements were performed at room temperature. The photoemission endstation is composed of a hemispherical analyzer SPECS PHOIBOS 150 and a FERRUM 3D spin detector \cite{Escher:2011}, based on very-low energy electron diffraction (VLEED).

The calculations were performed with the spin-polarized relativistic Korringa-Kohn-Rostoker (SPR-KKR) package \cite{Ebert:2011, Ebert:2017}, based on one-step photoemission theory within the spin-density matrix formalism \cite{Braun:2018}.

More details of the beamline, STARPES experiment and calculations are reported in Ref.~\cite{SOM:}.\\

In Fig.~\ref{fig:fig1} a spin-integrated characterization of the time evolution in WTe$_2$ is presented.
In Fig.~\ref{fig:fig1}~(b) four constant energy maps (CEMs) measured with $h\nu=35.65$~eV are shown: without pump and $100$~fs after a pump pulse, each for two energies, at $E_F$ and $400$~meV above it. The quality of the Fermi surfaces is limited by the energy resolution of the HHG pulse, but still good compared to literature \cite{Caputo:2018} and enough to distinguish hP and eP. Whereas without pump there is obviously no intensity above $E_F$, with the pump a clear signal from only eP is visible, confirming the expected shape of the band structure. The higher intensity at $E_F$ in the presence of the pump can be ascribed to thermal broadening.
Fig.~\ref{fig:fig1}~(c) shows the intensity integrated in the angular range shown by the blue rectangle in (b) and an energy range between $E_F$ and $500$~meV above it. The extracted time constant is of $\approx1$~ps, well compatible with literature \cite{Caputo:2018}. 

\begin{figure} [ht!]
	\centering
		\includegraphics[width=0.48\textwidth]{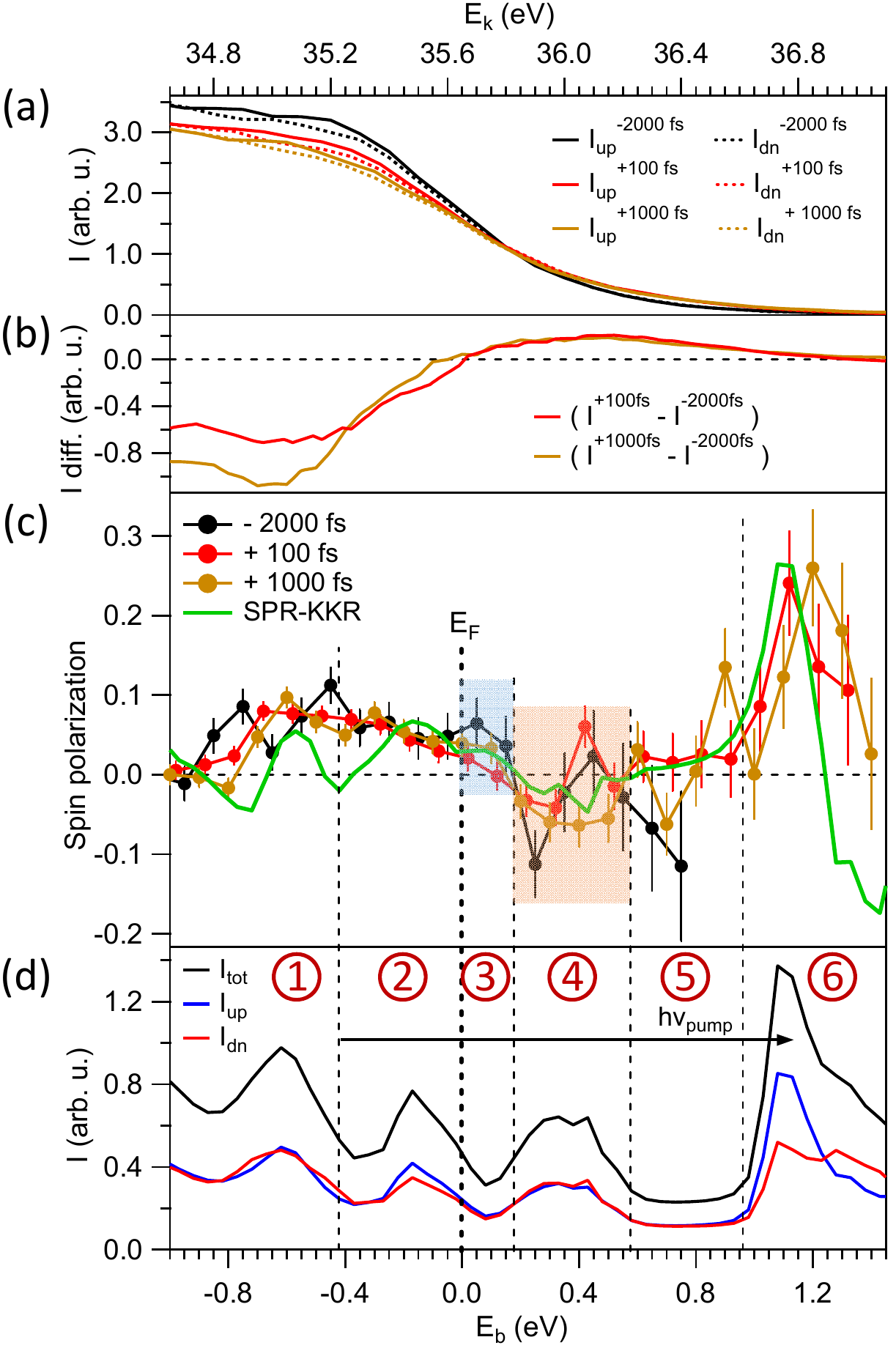}
	\caption{(a) $I^\uparrow$ (full) and $I^\downarrow$ (dotted) spin channels for three time delays. (b) Difference between total intensity after and before the pump. (c) Spin polarization along $\hat{\alpha}$. The calculated spin polarization with SPR-KKR in similar conditions as experimental ones is also shown. Highlighted areas are discussed in the text. (d) SPR-KKR calculations for $I^\uparrow$ and $I^\downarrow$ channels and their sum. The pump photon energy and the six energy regions considered in Fig.~\ref{fig:fig3} are shown.}
	\label{fig:fig2}
\end{figure}
Once the quality of the sample and reproducibility of its dynamics were confirmed, we performed the STARPES measurements.
The choosen spin quantization axis corresponds to a VLEED axis, $\hat{\alpha}=0.7\hat{x}+0.7\hat{y}+0.14\hat{z}$ in the sample geometry, as indicated in the $xy$ plane in Fig.~\ref{fig:fig1}~(a).
The angular acceptance of about $\pm2.5^\circ$ is indicated by the blue circle in Fig.~\ref{fig:fig1}~(a). This choice is due to the necessity of improving the count rate for the spin measurements, but still allows to distinguish the signal from a selected region in the Brillouin zone.
The UV is tuned to $35.65$~eV (23rd harmonic), which probes a cut near the $Z$ point along $\Gamma Z$ \cite{SOM:}.
In Fig.~\ref{fig:fig2}~(a) the spin up ($I^\uparrow$) and down ($I^\downarrow$) channels (full and dotted lines respectively) are shown for three different time delays, one before and two after the pump pulse.
The total intensity ($I^\uparrow+I^\downarrow$) differences between after and before the pump spectra are shown in Fig.~\ref{fig:fig2}~(b), illustrating the depletion of electrons below $E_F$ and the population above it.
The spin polarization along $\hat{\alpha}$ is calculated as $P=\frac{1}{S}\frac{I^\uparrow-I^\downarrow}{I^\uparrow+I^\downarrow}$ after constant background subtraction, with a Sherman function of $S=0.29$. Points with error bars larger than $10\%$ are discarded. The resulting curves for the three time delays are shown in Fig.~\ref{fig:fig2}~(c). The overall trend is a positive spin polarization for the occupied region, and a non-zero spin polarization for the unoccupied region with a up-down-up behaviour that crosses zero at about $+150$~meV and $+550$~meV. This trend is well confirmed by the spin polarization from SPR-KKR calculations also shown, performed with similar geometry and angular resolution as in the experiment.
In Fig.~\ref{fig:fig2}~(d) the corresponding $I^\uparrow$ and $I^\downarrow$ and their sum from SPR-KKR are shown.

\begin{figure} [h]
	\centering
		\includegraphics[width=0.48\textwidth]{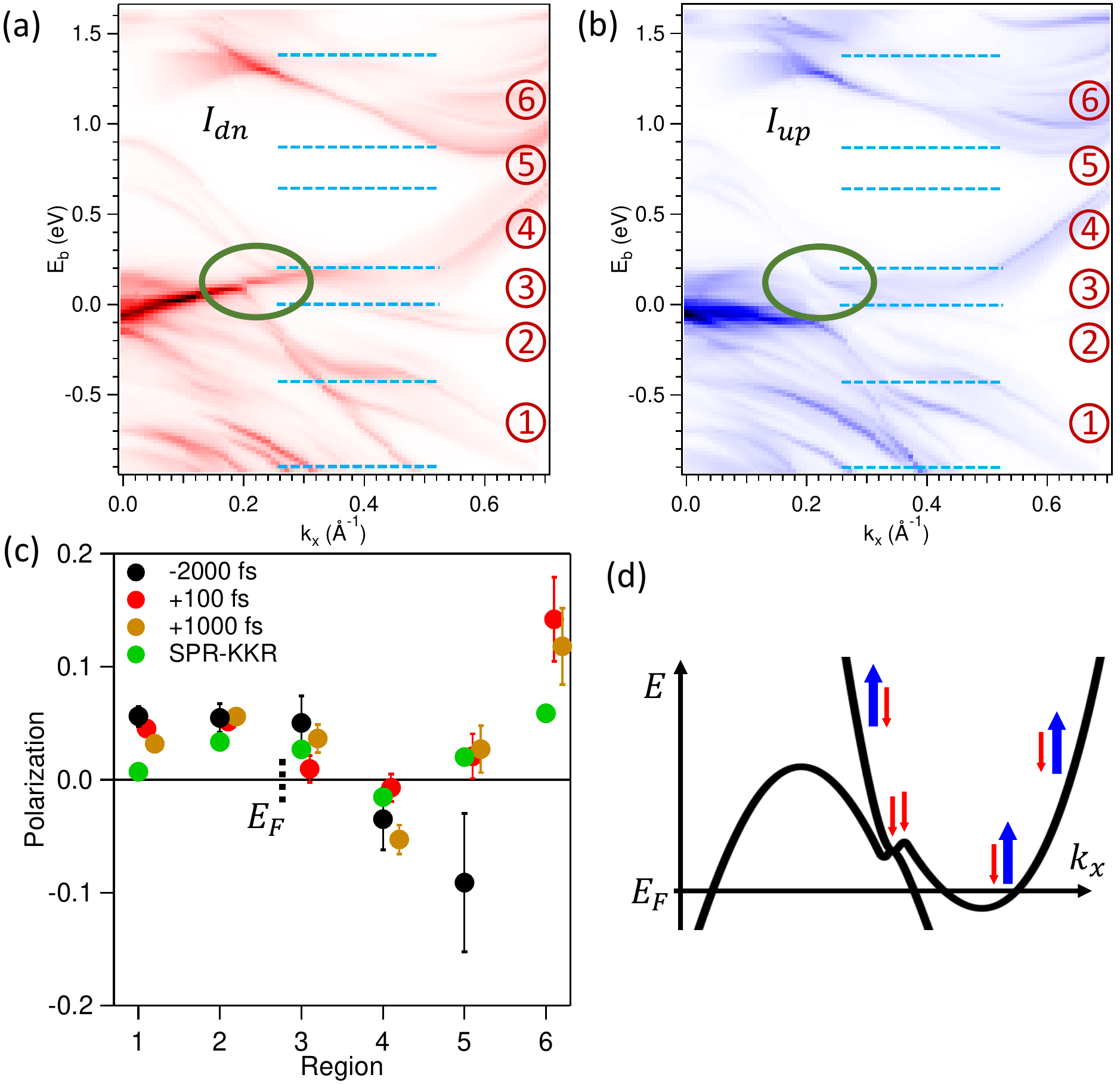}
	\caption{(a-b) SPR-KKR bandmap projected on $\hat{\alpha}$ along $\overline{\Gamma X}$ for $I^\downarrow$ and $I^\uparrow$ channels, respectively. The green circle highlights the WPs region. Dotted blue lines indicate the experimental angular resolution (note that $k_y\neq 0$ in the measurement). (c) Measured spin polarization at three time delays integrated in energy over each of the six regions of Fig.~\ref{fig:fig2}~(c). (d) Cartoon of the accumulation of spin down electrons above the WPs region.}
	\label{fig:fig3}
\end{figure}
With the support of the calculations, we can discuss several features of the measured spin polarization at the three time delays. We distinguish six regions, and their spin polarization can be better appreciated by averaging over the energy scale, as shown in Fig.~\ref{fig:fig3}~(c). The regions are also shown in the calculated band dispersion for $I^\downarrow$ and $I^\uparrow$ in Fig.~\ref{fig:fig3}~(a) and (b), respectively. The blue dotted lines represent the angular integration corresponding to the blue circle in Fig.~\ref{fig:fig1}~(a).
In region $5$, a gap in the density of states is found, therefore the measured small spin polarization is not very relevant.
In region $6$, on the other hand, new states become available, and the large peak of positive spin polarization measured for the two positive time delays is reproduced by the calculations.
Looking at the occupied part, electrons from region $1$ and below cannot reach region $6$ because of the pump photon energy ($1.55$~eV), and thus can only be excited into regions $3$ and $4$. Electrons from region $2$, on the other hand, will have the large density of states of region $6$ available. This observation helps to explain the differences of the measurements in region $1$ and $2$.
In region $2$ the spin polarization does not vary between the three time delays, while it does in region $1$. 
Thanks to the many states available and a matching spin polarization direction in region $6$ for electrons from region $2$, the amount of spin polarization here is not expected to vary drastically over time.
Instead, given the limited states available in region $3$ and $4$ and their opposite spin polarization, and because of the favorable depletion of electrons from region $2$, the positively polarized electrons in region $1$ are observed to change over time in a complex way.
This different behaviour for regions $1$ and $2$ is also observed in Fig.~\ref{fig:fig2}~(b), where the unexpected time evolution in region $1$ can be possibly explained by complex electron redistribution from other regions of the Brillouin zone.

In regions $3$ and $4$ above $E_F$, the measurement before the pump pulse is possible only because of the tail of the Fermi function and the large broadening of the spectra in Fig.~\ref{fig:fig2}~(a). 
The spin polarization sign matches the calculated one [see Fig.~\ref{fig:fig3}~(c)]. 
When the two time delays after the pump are considered, first a suppression and then a recover of the spin polarization is observed, both for the positive amount in region $3$ and the negative in region $4$ (blue and red areas in Fig.~\ref{fig:fig2}~(c), respectively). The effect is clear in Fig.~\ref{fig:fig3}~(c). 
According to the SPR-KKR calculations, region $3$ is where hP and eP are connected by SS and possibly by the WPs only for $I^\downarrow$ [green circles in Fig.~\ref{fig:fig3}~(a) and (b)], and region $4$ corresponds to the above bulk states. 
While after the short $+100$~fs time delay the spin polarization is suppressed, at a larger $+1000$~fs an accumulation of spin down is observed in region $4$. This can also be seen in the broad and homogeneous spin polarization peak at $+1000$~fs when compared to the other two time delays, as highlighted by the red area in Fig.~\ref{fig:fig2}~(c). 
The accumulation of spin above the energy where spin-polarized WPs are expected is an indication for the presence of Weyl (or slightly gapped Weyl-like) cones. 
A WP is expected to act as a bottleneck which selectively decelerates the evacuation of a spin channel, in a similar way as the (spin-independent) slower evacuation above the (spin-degenerate) Dirac point of graphene \cite{Johannsen:2015}.
A cartoon of the accumulation of down electrons in region $4$ is shown in Fig.~\ref{fig:fig3}~(d). The picture does not consider whether there are zero, two or four couples of WPs, but that an overall cone-like and spin polarized electronic structure is present. 

\begin{figure} [h]
	\centering
		\includegraphics[width=0.48\textwidth]{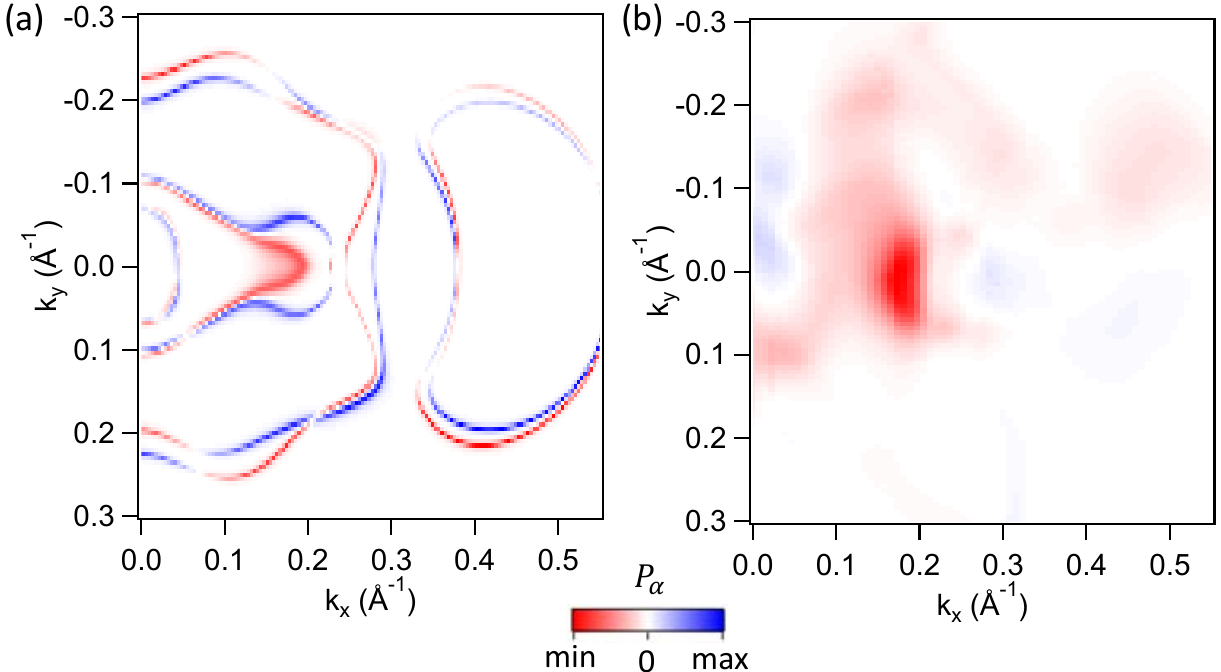}
	\caption{(a) Spin-resolved initial state Bloch spectral function at the $k_z$ point probed by $h\nu=35.65$~eV \cite{SOM:} and (b) corresponding SPR-KKR photoemission calculation, for a CEM at $55$~meV above $E_F$. The spin quantization axis is $\hat{\alpha}$.}
	\label{fig:fig4}
\end{figure}
In order to have a better insight in the spin texture, in Fig.~\ref{fig:fig4} a comparison is shown between spin-resolved ground state Bloch spectral function (a) and SPR-KKR photoemission calculation (b), for a CEM at $55$~meV above $E_F$. 
The spin quantization axis is $\hat{\alpha}$ 
and the $k_z$ position is the one probed by $h\nu=35.65$~eV, which is far from $\Gamma$ where WPs should occur \cite{Soluyanov:2015}. 
While the spin polarization in the WPs region for the ground state changes sign many times, its complexity is washed out when photoemission is performed. 
Furthermore, only small quantitative differences are found when comparing with the $Z$ or $\Gamma$ points, as reported in Ref.~\cite{SOM:}. Therefore, it is clear that in a photoemission experiment on WTe$_2$ it is not relevant to consider the precise topological classification, but rather to study the overall peculiar spin dynamics.\\

In this paper, we presented the first experimental STARPES study with an HHG source performed on WTe$_2$.
STARPES measurements are certainly demanding and time-consuming, but thanks to increasingly better control of HHG sources they can become a viable complementary alternative to synchrotron-based studies for different purposes \cite{SOM:}.
A parallel theoretical effort to explore not only charge but also spin dynamics in out-of-equilibrium scenarios is required \cite{Battiato:2018}.
Indeed, a clear limitation of our work is that supporting calculations are performed at equilibrium. However, their remarkable agreement with the experimental overall trend proves that in our conditions the spin polarization qualitative properties are not heavily affected when out-of-equilibrium.

Our STARPES results on WTe$_2$ show a non-zero spin polarization above $E_F$, crossing zero at $150$~meV and $550$~meV above it.
Given the instability of the topological classification and number of WPs upon small variations of structural parameters \cite{Russmann:2018}, and eventually their extreme vicinity \cite{Soluyanov:2015}, we find not useful to try to unambiguously label WTe$_2$ other than being near to a topological phase transition.
Instead, we studied the collective behaviour of spin polarization upon excitation, and we observed a quick suppression at $100$~fs after pump and subsequent accumulation of electron spins at $1000$~fs in the region above the expected WPs.
This suggests a spin-selective bottleneck effect, which is in line with the expected electronic structure above $E_F$. 
Whether the spin polarized cone-like structure is gapless and thus is hosting topological low energy excitations dubbed as Weyl fermions, or the structure is just a gapped precondition for it and the electron excitation is trivial, it is irrelevant for the overall spin dynamics. Instead, it is the peculiar spin-polarized, cone-like electronic structure of WTe$_2$ that is useful to explain its anomalous transport properties.\\

We dedicate this work to the memory of our colleague and friend B. Carré. We gratefully acknowledge discussions with L. Nicolaï. 

M. F. aknowledges support by the Swiss National Science Foundation project No. P2ELP2\_181877. 
The laser system and the experimental setup are supported by the French "`Investments for the Future"' of the Agence Nationale pour la Recherche, contracts No. 11-EQPX0005-ATTOLAB and 11-EQPX0034-PATRIMEX. 
The laser system is also supported by the Scientific Cooperation Foundation of Paris-Saclay University through the funding of the OPT2X research project (Lidex 2014) 
and by the Ile de France region through the Pulse-X project.
J. S. and J. M. would like to thank CEDAMNF project financed by the Ministry of Education, Youth and Sports of Czech Repuplic, Project No. CZ.02.1.01/0.0/0.0/15.003/0000358.

M. F. and J. S. contributed equally to this work.
\footnotesize

\begin{thebibliography}{40}%
\makeatletter
\providecommand \@ifxundefined [1]{%
 \@ifx{#1\undefined}
}%
\providecommand \@ifnum [1]{%
 \ifnum #1\expandafter \@firstoftwo
 \else \expandafter \@secondoftwo
 \fi
}%
\providecommand \@ifx [1]{%
 \ifx #1\expandafter \@firstoftwo
 \else \expandafter \@secondoftwo
 \fi
}%
\providecommand \natexlab [1]{#1}%
\providecommand \enquote  [1]{``#1''}%
\providecommand \bibnamefont  [1]{#1}%
\providecommand \bibfnamefont [1]{#1}%
\providecommand \citenamefont [1]{#1}%
\providecommand \href@noop [0]{\@secondoftwo}%
\providecommand \href [0]{\begingroup \@sanitize@url \@href}%
\providecommand \@href[1]{\@@startlink{#1}\@@href}%
\providecommand \@@href[1]{\endgroup#1\@@endlink}%
\providecommand \@sanitize@url [0]{\catcode `\\12\catcode `\$12\catcode
  `\&12\catcode `\#12\catcode `\^12\catcode `\_12\catcode `\%12\relax}%
\providecommand \@@startlink[1]{}%
\providecommand \@@endlink[0]{}%
\providecommand \url  [0]{\begingroup\@sanitize@url \@url }%
\providecommand \@url [1]{\endgroup\@href {#1}{\urlprefix }}%
\providecommand \urlprefix  [0]{URL }%
\providecommand \Eprint [0]{\href }%
\providecommand \doibase [0]{http://dx.doi.org/}%
\providecommand \selectlanguage [0]{\@gobble}%
\providecommand \bibinfo  [0]{\@secondoftwo}%
\providecommand \bibfield  [0]{\@secondoftwo}%
\providecommand \translation [1]{[#1]}%
\providecommand \BibitemOpen [0]{}%
\providecommand \bibitemStop [0]{}%
\providecommand \bibitemNoStop [0]{.\EOS\space}%
\providecommand \EOS [0]{\spacefactor3000\relax}%
\providecommand \BibitemShut  [1]{\csname bibitem#1\endcsname}%
\let\auto@bib@innerbib\@empty
\bibitem [{\citenamefont {Ali}\ \emph {et~al.}(2014)\citenamefont {Ali},
  \citenamefont {Xiong}, \citenamefont {Flynn}, \citenamefont {Tao},
  \citenamefont {Gibson}, \citenamefont {Schoop}, \citenamefont {Liang},
  \citenamefont {Haldolaarachchige}, \citenamefont {Hirschberger},
  \citenamefont {Ong},\ and\ \citenamefont {Cava}}]{Ali:2014}%
  \BibitemOpen
  \bibfield  {author} {\bibinfo {author} {\bibfnamefont {M.~N.}\ \bibnamefont
  {Ali}}, \bibinfo {author} {\bibfnamefont {J.}~\bibnamefont {Xiong}}, \bibinfo
  {author} {\bibfnamefont {S.}~\bibnamefont {Flynn}}, \bibinfo {author}
  {\bibfnamefont {J.}~\bibnamefont {Tao}}, \bibinfo {author} {\bibfnamefont
  {Q.~D.}\ \bibnamefont {Gibson}}, \bibinfo {author} {\bibfnamefont {L.~M.}\
  \bibnamefont {Schoop}}, \bibinfo {author} {\bibfnamefont {T.}~\bibnamefont
  {Liang}}, \bibinfo {author} {\bibfnamefont {N.}~\bibnamefont
  {Haldolaarachchige}}, \bibinfo {author} {\bibfnamefont {M.}~\bibnamefont
  {Hirschberger}}, \bibinfo {author} {\bibfnamefont {N.~P.}\ \bibnamefont
  {Ong}}, \ and\ \bibinfo {author} {\bibfnamefont {R.~J.}\ \bibnamefont
  {Cava}},\ }\href {https://doi.org/10.1038/nature13763
  http://10.0.4.14/nature13763} {\bibfield  {journal} {\bibinfo  {journal}
  {Nature}\ }\textbf {\bibinfo {volume} {514}},\ \bibinfo {pages} {205}
  (\bibinfo {year} {2014})}\BibitemShut {NoStop}%
\bibitem [{\citenamefont {Pan}\ \emph {et~al.}(2015)\citenamefont {Pan},
  \citenamefont {Chen}, \citenamefont {Liu}, \citenamefont {Feng},
  \citenamefont {Wei}, \citenamefont {Zhou}, \citenamefont {Chi}, \citenamefont
  {Pi}, \citenamefont {Yen}, \citenamefont {Song}, \citenamefont {Wan},
  \citenamefont {Yang}, \citenamefont {Wang}, \citenamefont {Wang},\ and\
  \citenamefont {Zhang}}]{Pan:2015}%
  \BibitemOpen
  \bibfield  {author} {\bibinfo {author} {\bibfnamefont {X.-C.}\ \bibnamefont
  {Pan}}, \bibinfo {author} {\bibfnamefont {X.}~\bibnamefont {Chen}}, \bibinfo
  {author} {\bibfnamefont {H.}~\bibnamefont {Liu}}, \bibinfo {author}
  {\bibfnamefont {Y.}~\bibnamefont {Feng}}, \bibinfo {author} {\bibfnamefont
  {Z.}~\bibnamefont {Wei}}, \bibinfo {author} {\bibfnamefont {Y.}~\bibnamefont
  {Zhou}}, \bibinfo {author} {\bibfnamefont {Z.}~\bibnamefont {Chi}}, \bibinfo
  {author} {\bibfnamefont {L.}~\bibnamefont {Pi}}, \bibinfo {author}
  {\bibfnamefont {F.}~\bibnamefont {Yen}}, \bibinfo {author} {\bibfnamefont
  {F.}~\bibnamefont {Song}}, \bibinfo {author} {\bibfnamefont {X.}~\bibnamefont
  {Wan}}, \bibinfo {author} {\bibfnamefont {Z.}~\bibnamefont {Yang}}, \bibinfo
  {author} {\bibfnamefont {B.}~\bibnamefont {Wang}}, \bibinfo {author}
  {\bibfnamefont {G.}~\bibnamefont {Wang}}, \ and\ \bibinfo {author}
  {\bibfnamefont {Y.}~\bibnamefont {Zhang}},\ }\href {\doibase
  10.1038/ncomms8805} {\bibfield  {journal} {\bibinfo  {journal} {Nature
  Communications}\ }\textbf {\bibinfo {volume} {6}},\ \bibinfo {pages} {7805}
  (\bibinfo {year} {2015})}\BibitemShut {NoStop}%
\bibitem [{\citenamefont {Soluyanov}\ \emph {et~al.}(2015)\citenamefont
  {Soluyanov}, \citenamefont {Gresch}, \citenamefont {Wang}, \citenamefont
  {Wu}, \citenamefont {Troyer}, \citenamefont {Dai},\ and\ \citenamefont
  {Bernevig}}]{Soluyanov:2015}%
  \BibitemOpen
  \bibfield  {author} {\bibinfo {author} {\bibfnamefont {A.~A.}\ \bibnamefont
  {Soluyanov}}, \bibinfo {author} {\bibfnamefont {D.}~\bibnamefont {Gresch}},
  \bibinfo {author} {\bibfnamefont {Z.}~\bibnamefont {Wang}}, \bibinfo {author}
  {\bibfnamefont {Q.}~\bibnamefont {Wu}}, \bibinfo {author} {\bibfnamefont
  {M.}~\bibnamefont {Troyer}}, \bibinfo {author} {\bibfnamefont
  {X.}~\bibnamefont {Dai}}, \ and\ \bibinfo {author} {\bibfnamefont {B.~A.}\
  \bibnamefont {Bernevig}},\ }\href {https://doi.org/10.1038/nature15768
  http://10.0.4.14/nature15768
  https://www.nature.com/articles/nature15768{\#}supplementary-information}
  {\bibfield  {journal} {\bibinfo  {journal} {Nature}\ }\textbf {\bibinfo
  {volume} {527}},\ \bibinfo {pages} {495} (\bibinfo {year}
  {2015})}\BibitemShut {NoStop}%
\bibitem [{\citenamefont {Abrikosov}\ and\ \citenamefont
  {Beneslavskii}(1971)}]{Abrikosov:1971}%
  \BibitemOpen
  \bibfield  {author} {\bibinfo {author} {\bibfnamefont {A.~A.}\ \bibnamefont
  {Abrikosov}}\ and\ \bibinfo {author} {\bibfnamefont {S.~D.}\ \bibnamefont
  {Beneslavskii}},\ }\href@noop {} {\bibfield  {journal} {\bibinfo  {journal}
  {Journal of Experimental and Theoretical Physics}\ }\textbf {\bibinfo
  {volume} {32}},\ \bibinfo {pages} {699} (\bibinfo {year} {1971})}\BibitemShut
  {NoStop}%
\bibitem [{\citenamefont {Tamai}\ \emph {et~al.}(2016)\citenamefont {Tamai},
  \citenamefont {Wu}, \citenamefont {Cucchi}, \citenamefont {Bruno},
  \citenamefont {Ricc{\`{o}}}, \citenamefont {Kim}, \citenamefont {Hoesch},
  \citenamefont {Barreteau}, \citenamefont {Giannini}, \citenamefont {Besnard},
  \citenamefont {Soluyanov},\ and\ \citenamefont {Baumberger}}]{Tamai:2016}%
  \BibitemOpen
  \bibfield  {author} {\bibinfo {author} {\bibfnamefont {A.}~\bibnamefont
  {Tamai}}, \bibinfo {author} {\bibfnamefont {Q.~S.}\ \bibnamefont {Wu}},
  \bibinfo {author} {\bibfnamefont {I.}~\bibnamefont {Cucchi}}, \bibinfo
  {author} {\bibfnamefont {F.~Y.}\ \bibnamefont {Bruno}}, \bibinfo {author}
  {\bibfnamefont {S.}~\bibnamefont {Ricc{\`{o}}}}, \bibinfo {author}
  {\bibfnamefont {T.~K.}\ \bibnamefont {Kim}}, \bibinfo {author} {\bibfnamefont
  {M.}~\bibnamefont {Hoesch}}, \bibinfo {author} {\bibfnamefont
  {C.}~\bibnamefont {Barreteau}}, \bibinfo {author} {\bibfnamefont
  {E.}~\bibnamefont {Giannini}}, \bibinfo {author} {\bibfnamefont
  {C.}~\bibnamefont {Besnard}}, \bibinfo {author} {\bibfnamefont {A.~A.}\
  \bibnamefont {Soluyanov}}, \ and\ \bibinfo {author} {\bibfnamefont
  {F.}~\bibnamefont {Baumberger}},\ }\href {\doibase 10.1103/PhysRevX.6.031021}
  {\bibfield  {journal} {\bibinfo  {journal} {Phys. Rev. X}\ }\textbf {\bibinfo
  {volume} {6}},\ \bibinfo {pages} {31021} (\bibinfo {year}
  {2016})}\BibitemShut {NoStop}%
\bibitem [{\citenamefont {Weber}\ \emph {et~al.}(2018)\citenamefont {Weber},
  \citenamefont {R{\"{u}}ssmann}, \citenamefont {Xu}, \citenamefont {Muff},
  \citenamefont {Fanciulli}, \citenamefont {Magrez}, \citenamefont {Bugnon},
  \citenamefont {Berger}, \citenamefont {Plumb}, \citenamefont {Shi},
  \citenamefont {Bl{\"{u}}gel}, \citenamefont {Mavropoulos},\ and\
  \citenamefont {Dil}}]{Weber:2018}%
  \BibitemOpen
  \bibfield  {author} {\bibinfo {author} {\bibfnamefont {A.~P.}\ \bibnamefont
  {Weber}}, \bibinfo {author} {\bibfnamefont {P.}~\bibnamefont
  {R{\"{u}}ssmann}}, \bibinfo {author} {\bibfnamefont {N.}~\bibnamefont {Xu}},
  \bibinfo {author} {\bibfnamefont {S.}~\bibnamefont {Muff}}, \bibinfo {author}
  {\bibfnamefont {M.}~\bibnamefont {Fanciulli}}, \bibinfo {author}
  {\bibfnamefont {A.}~\bibnamefont {Magrez}}, \bibinfo {author} {\bibfnamefont
  {P.}~\bibnamefont {Bugnon}}, \bibinfo {author} {\bibfnamefont
  {H.}~\bibnamefont {Berger}}, \bibinfo {author} {\bibfnamefont {N.~C.}\
  \bibnamefont {Plumb}}, \bibinfo {author} {\bibfnamefont {M.}~\bibnamefont
  {Shi}}, \bibinfo {author} {\bibfnamefont {S.}~\bibnamefont {Bl{\"{u}}gel}},
  \bibinfo {author} {\bibfnamefont {P.}~\bibnamefont {Mavropoulos}}, \ and\
  \bibinfo {author} {\bibfnamefont {J.~H.}\ \bibnamefont {Dil}},\ }\href
  {\doibase 10.1103/PhysRevLett.121.156401} {\bibfield  {journal} {\bibinfo
  {journal} {Phys. Rev. Lett.}\ }\textbf {\bibinfo {volume} {121}},\ \bibinfo
  {pages} {156401} (\bibinfo {year} {2018})}\BibitemShut {NoStop}%
\bibitem [{\citenamefont {R{\"{u}}ssmann}\ \emph {et~al.}(2018)\citenamefont
  {R{\"{u}}ssmann}, \citenamefont {Weber}, \citenamefont {Glott}, \citenamefont
  {Xu}, \citenamefont {Fanciulli}, \citenamefont {Muff}, \citenamefont
  {Magrez}, \citenamefont {Bugnon}, \citenamefont {Berger}, \citenamefont
  {Bode}, \citenamefont {Dil}, \citenamefont {Bl{\"{u}}gel}, \citenamefont
  {Mavropoulos},\ and\ \citenamefont {Sessi}}]{Russmann:2018}%
  \BibitemOpen
  \bibfield  {author} {\bibinfo {author} {\bibfnamefont {P.}~\bibnamefont
  {R{\"{u}}ssmann}}, \bibinfo {author} {\bibfnamefont {A.~P.}\ \bibnamefont
  {Weber}}, \bibinfo {author} {\bibfnamefont {F.}~\bibnamefont {Glott}},
  \bibinfo {author} {\bibfnamefont {N.}~\bibnamefont {Xu}}, \bibinfo {author}
  {\bibfnamefont {M.}~\bibnamefont {Fanciulli}}, \bibinfo {author}
  {\bibfnamefont {S.}~\bibnamefont {Muff}}, \bibinfo {author} {\bibfnamefont
  {A.}~\bibnamefont {Magrez}}, \bibinfo {author} {\bibfnamefont
  {P.}~\bibnamefont {Bugnon}}, \bibinfo {author} {\bibfnamefont
  {H.}~\bibnamefont {Berger}}, \bibinfo {author} {\bibfnamefont
  {M.}~\bibnamefont {Bode}}, \bibinfo {author} {\bibfnamefont {J.~H.}\
  \bibnamefont {Dil}}, \bibinfo {author} {\bibfnamefont {S.}~\bibnamefont
  {Bl{\"{u}}gel}}, \bibinfo {author} {\bibfnamefont {P.}~\bibnamefont
  {Mavropoulos}}, \ and\ \bibinfo {author} {\bibfnamefont {P.}~\bibnamefont
  {Sessi}},\ }\href {\doibase 10.1103/PhysRevB.97.075106} {\bibfield  {journal}
  {\bibinfo  {journal} {Phys. Rev. B}\ }\textbf {\bibinfo {volume} {97}},\
  \bibinfo {pages} {75106} (\bibinfo {year} {2018})}\BibitemShut {NoStop}%
\bibitem [{\citenamefont {Lv}\ \emph {et~al.}(2017)\citenamefont {Lv},
  \citenamefont {Li}, \citenamefont {Zhang}, \citenamefont {Deng},
  \citenamefont {Yao}, \citenamefont {Chen}, \citenamefont {Zhou},
  \citenamefont {Zhang}, \citenamefont {Lu}, \citenamefont {Zhang},
  \citenamefont {Tian}, \citenamefont {Sheng},\ and\ \citenamefont
  {Chen}}]{Lv:2017}%
  \BibitemOpen
  \bibfield  {author} {\bibinfo {author} {\bibfnamefont {Y.-Y.}\ \bibnamefont
  {Lv}}, \bibinfo {author} {\bibfnamefont {X.}~\bibnamefont {Li}}, \bibinfo
  {author} {\bibfnamefont {B.-B.}\ \bibnamefont {Zhang}}, \bibinfo {author}
  {\bibfnamefont {W.~Y.}\ \bibnamefont {Deng}}, \bibinfo {author}
  {\bibfnamefont {S.-H.}\ \bibnamefont {Yao}}, \bibinfo {author} {\bibfnamefont
  {Y.~B.}\ \bibnamefont {Chen}}, \bibinfo {author} {\bibfnamefont
  {J.}~\bibnamefont {Zhou}}, \bibinfo {author} {\bibfnamefont {S.-T.}\
  \bibnamefont {Zhang}}, \bibinfo {author} {\bibfnamefont {M.-H.}\ \bibnamefont
  {Lu}}, \bibinfo {author} {\bibfnamefont {L.}~\bibnamefont {Zhang}}, \bibinfo
  {author} {\bibfnamefont {M.}~\bibnamefont {Tian}}, \bibinfo {author}
  {\bibfnamefont {L.}~\bibnamefont {Sheng}}, \ and\ \bibinfo {author}
  {\bibfnamefont {Y.-F.}\ \bibnamefont {Chen}},\ }\href {\doibase
  10.1103/PhysRevLett.118.096603} {\bibfield  {journal} {\bibinfo  {journal}
  {Phys. Rev. Lett.}\ }\textbf {\bibinfo {volume} {118}},\ \bibinfo {pages}
  {96603} (\bibinfo {year} {2017})}\BibitemShut {NoStop}%
\bibitem [{\citenamefont {Zhang}\ \emph {et~al.}(2017)\citenamefont {Zhang},
  \citenamefont {Wu}, \citenamefont {Zhang}, \citenamefont {Cheong},
  \citenamefont {Soluyanov},\ and\ \citenamefont {Wu}}]{Zhang:2017}%
  \BibitemOpen
  \bibfield  {author} {\bibinfo {author} {\bibfnamefont {W.}~\bibnamefont
  {Zhang}}, \bibinfo {author} {\bibfnamefont {Q.}~\bibnamefont {Wu}}, \bibinfo
  {author} {\bibfnamefont {L.}~\bibnamefont {Zhang}}, \bibinfo {author}
  {\bibfnamefont {S.-W.}\ \bibnamefont {Cheong}}, \bibinfo {author}
  {\bibfnamefont {A.~A.}\ \bibnamefont {Soluyanov}}, \ and\ \bibinfo {author}
  {\bibfnamefont {W.}~\bibnamefont {Wu}},\ }\href {\doibase
  10.1103/PhysRevB.96.165125} {\bibfield  {journal} {\bibinfo  {journal} {Phys.
  Rev. B}\ }\textbf {\bibinfo {volume} {96}},\ \bibinfo {pages} {165125}
  (\bibinfo {year} {2017})}\BibitemShut {NoStop}%
\bibitem [{SOM()}]{SOM:}%
  \BibitemOpen
  \href@noop {} {{\bibinfo {title} {{See \textit{Supplemental Material} at
  [URL], which includes Ref.~\cite{Norman:1998}, for a comment on Fermi arcs, for details of the beamline, measurements and calculations, and for further results from calculations}}}\ }\BibitemShut {NoStop}%
\bibitem [{\citenamefont {Norman}\ \emph {et~al.}(1998)\citenamefont {Norman},
  \citenamefont {Ding}, \citenamefont {Randeria}, \citenamefont {Campuzano},
  \citenamefont {Yokoya}, \citenamefont {Takeuchi}, \citenamefont {Takahashi},
  \citenamefont {Mochiku}, \citenamefont {Kadowaki}, \citenamefont
  {Guptasarma},\ and\ \citenamefont {Hinks}}]{Norman:1998}%
  \BibitemOpen
  \bibfield  {author} {\bibinfo {author} {\bibfnamefont {M.~R.}\ \bibnamefont
  {Norman}}, \bibinfo {author} {\bibfnamefont {H.}~\bibnamefont {Ding}},
  \bibinfo {author} {\bibfnamefont {M.}~\bibnamefont {Randeria}}, \bibinfo
  {author} {\bibfnamefont {J.~C.}\ \bibnamefont {Campuzano}}, \bibinfo {author}
  {\bibfnamefont {T.}~\bibnamefont {Yokoya}}, \bibinfo {author} {\bibfnamefont
  {T.}~\bibnamefont {Takeuchi}}, \bibinfo {author} {\bibfnamefont
  {T.}~\bibnamefont {Takahashi}}, \bibinfo {author} {\bibfnamefont
  {T.}~\bibnamefont {Mochiku}}, \bibinfo {author} {\bibfnamefont
  {K.}~\bibnamefont {Kadowaki}}, \bibinfo {author} {\bibfnamefont
  {P.}~\bibnamefont {Guptasarma}}, \ and\ \bibinfo {author} {\bibfnamefont
  {D.~G.}\ \bibnamefont {Hinks}},\ }\href {https://doi.org/10.1038/32366
  http://10.0.4.14/32366} {\bibfield  {journal} {\bibinfo  {journal} {Nature}\
  }\textbf {\bibinfo {volume} {392}},\ \bibinfo {pages} {157} (\bibinfo {year}
  {1998})}\BibitemShut {NoStop}%
\bibitem [{\citenamefont {Wang}\ \emph {et~al.}(2016)\citenamefont {Wang},
  \citenamefont {Zhang}, \citenamefont {Huang}, \citenamefont {Nie},
  \citenamefont {Liu}, \citenamefont {Liang}, \citenamefont {Zhang},
  \citenamefont {Shen}, \citenamefont {Liu}, \citenamefont {Hu}, \citenamefont
  {Ding}, \citenamefont {Liu}, \citenamefont {Hu}, \citenamefont {He},
  \citenamefont {Zhao}, \citenamefont {Yu}, \citenamefont {Hu}, \citenamefont
  {Wei}, \citenamefont {Mao}, \citenamefont {Shi}, \citenamefont {Jia},
  \citenamefont {Zhang}, \citenamefont {Zhang}, \citenamefont {Yang},
  \citenamefont {Wang}, \citenamefont {Peng}, \citenamefont {Weng},
  \citenamefont {Dai}, \citenamefont {Fang}, \citenamefont {Xu}, \citenamefont
  {Chen},\ and\ \citenamefont {Zhou}}]{Wang:2016}%
  \BibitemOpen
  \bibfield  {author} {\bibinfo {author} {\bibfnamefont {C.}~\bibnamefont
  {Wang}}, \bibinfo {author} {\bibfnamefont {Y.}~\bibnamefont {Zhang}},
  \bibinfo {author} {\bibfnamefont {J.}~\bibnamefont {Huang}}, \bibinfo
  {author} {\bibfnamefont {S.}~\bibnamefont {Nie}}, \bibinfo {author}
  {\bibfnamefont {G.}~\bibnamefont {Liu}}, \bibinfo {author} {\bibfnamefont
  {A.}~\bibnamefont {Liang}}, \bibinfo {author} {\bibfnamefont
  {Y.}~\bibnamefont {Zhang}}, \bibinfo {author} {\bibfnamefont
  {B.}~\bibnamefont {Shen}}, \bibinfo {author} {\bibfnamefont {J.}~\bibnamefont
  {Liu}}, \bibinfo {author} {\bibfnamefont {C.}~\bibnamefont {Hu}}, \bibinfo
  {author} {\bibfnamefont {Y.}~\bibnamefont {Ding}}, \bibinfo {author}
  {\bibfnamefont {D.}~\bibnamefont {Liu}}, \bibinfo {author} {\bibfnamefont
  {Y.}~\bibnamefont {Hu}}, \bibinfo {author} {\bibfnamefont {S.}~\bibnamefont
  {He}}, \bibinfo {author} {\bibfnamefont {L.}~\bibnamefont {Zhao}}, \bibinfo
  {author} {\bibfnamefont {L.}~\bibnamefont {Yu}}, \bibinfo {author}
  {\bibfnamefont {J.}~\bibnamefont {Hu}}, \bibinfo {author} {\bibfnamefont
  {J.}~\bibnamefont {Wei}}, \bibinfo {author} {\bibfnamefont {Z.}~\bibnamefont
  {Mao}}, \bibinfo {author} {\bibfnamefont {Y.}~\bibnamefont {Shi}}, \bibinfo
  {author} {\bibfnamefont {X.}~\bibnamefont {Jia}}, \bibinfo {author}
  {\bibfnamefont {F.}~\bibnamefont {Zhang}}, \bibinfo {author} {\bibfnamefont
  {S.}~\bibnamefont {Zhang}}, \bibinfo {author} {\bibfnamefont
  {F.}~\bibnamefont {Yang}}, \bibinfo {author} {\bibfnamefont {Z.}~\bibnamefont
  {Wang}}, \bibinfo {author} {\bibfnamefont {Q.}~\bibnamefont {Peng}}, \bibinfo
  {author} {\bibfnamefont {H.}~\bibnamefont {Weng}}, \bibinfo {author}
  {\bibfnamefont {X.}~\bibnamefont {Dai}}, \bibinfo {author} {\bibfnamefont
  {Z.}~\bibnamefont {Fang}}, \bibinfo {author} {\bibfnamefont {Z.}~\bibnamefont
  {Xu}}, \bibinfo {author} {\bibfnamefont {C.}~\bibnamefont {Chen}}, \ and\
  \bibinfo {author} {\bibfnamefont {X.~J.}\ \bibnamefont {Zhou}},\ }\href
  {\doibase 10.1103/PhysRevB.94.241119} {\bibfield  {journal} {\bibinfo
  {journal} {Phys. Rev. B}\ }\textbf {\bibinfo {volume} {94}},\ \bibinfo
  {pages} {241119(R)} (\bibinfo {year} {2016})}\BibitemShut {NoStop}%
\bibitem [{\citenamefont {Das}\ \emph {et~al.}(2016)\citenamefont {Das},
  \citenamefont {{Di Sante}}, \citenamefont {Vobornik}, \citenamefont {Fujii},
  \citenamefont {Okuda}, \citenamefont {Bruyer}, \citenamefont {Gyenis},
  \citenamefont {Feldman}, \citenamefont {Tao}, \citenamefont {Ciancio},
  \citenamefont {Rossi}, \citenamefont {Ali}, \citenamefont {Picozzi},
  \citenamefont {Yadzani}, \citenamefont {Panaccione},\ and\ \citenamefont
  {Cava}}]{Das:2016}%
  \BibitemOpen
  \bibfield  {author} {\bibinfo {author} {\bibfnamefont {P.~K.}\ \bibnamefont
  {Das}}, \bibinfo {author} {\bibfnamefont {D.}~\bibnamefont {{Di Sante}}},
  \bibinfo {author} {\bibfnamefont {I.}~\bibnamefont {Vobornik}}, \bibinfo
  {author} {\bibfnamefont {J.}~\bibnamefont {Fujii}}, \bibinfo {author}
  {\bibfnamefont {T.}~\bibnamefont {Okuda}}, \bibinfo {author} {\bibfnamefont
  {E.}~\bibnamefont {Bruyer}}, \bibinfo {author} {\bibfnamefont
  {A.}~\bibnamefont {Gyenis}}, \bibinfo {author} {\bibfnamefont {B.~E.}\
  \bibnamefont {Feldman}}, \bibinfo {author} {\bibfnamefont {J.}~\bibnamefont
  {Tao}}, \bibinfo {author} {\bibfnamefont {R.}~\bibnamefont {Ciancio}},
  \bibinfo {author} {\bibfnamefont {G.}~\bibnamefont {Rossi}}, \bibinfo
  {author} {\bibfnamefont {M.~N.}\ \bibnamefont {Ali}}, \bibinfo {author}
  {\bibfnamefont {S.}~\bibnamefont {Picozzi}}, \bibinfo {author} {\bibfnamefont
  {A.}~\bibnamefont {Yadzani}}, \bibinfo {author} {\bibfnamefont
  {G.}~\bibnamefont {Panaccione}}, \ and\ \bibinfo {author} {\bibfnamefont
  {R.~J.}\ \bibnamefont {Cava}},\ }\href {https://doi.org/10.1038/ncomms10847
  http://10.0.4.14/ncomms10847
  https://www.nature.com/articles/ncomms10847{\#}supplementary-information}
  {\bibfield  {journal} {\bibinfo  {journal} {Nature Communications}\ }\textbf
  {\bibinfo {volume} {7}},\ \bibinfo {pages} {10847} (\bibinfo {year}
  {2016})}\BibitemShut {NoStop}%
\bibitem [{\citenamefont {Wu}\ \emph {et~al.}(2016)\citenamefont {Wu},
  \citenamefont {Mou}, \citenamefont {Jo}, \citenamefont {Sun}, \citenamefont
  {Huang}, \citenamefont {Bud'ko}, \citenamefont {Canfield},\ and\
  \citenamefont {Kaminski}}]{Wu:2016}%
  \BibitemOpen
  \bibfield  {author} {\bibinfo {author} {\bibfnamefont {Y.}~\bibnamefont
  {Wu}}, \bibinfo {author} {\bibfnamefont {D.}~\bibnamefont {Mou}}, \bibinfo
  {author} {\bibfnamefont {N.~H.}\ \bibnamefont {Jo}}, \bibinfo {author}
  {\bibfnamefont {K.}~\bibnamefont {Sun}}, \bibinfo {author} {\bibfnamefont
  {L.}~\bibnamefont {Huang}}, \bibinfo {author} {\bibfnamefont {S.~L.}\
  \bibnamefont {Bud'ko}}, \bibinfo {author} {\bibfnamefont {P.~C.}\
  \bibnamefont {Canfield}}, \ and\ \bibinfo {author} {\bibfnamefont
  {A.}~\bibnamefont {Kaminski}},\ }\href {\doibase 10.1103/PhysRevB.94.121113}
  {\bibfield  {journal} {\bibinfo  {journal} {Phys. Rev. B}\ }\textbf {\bibinfo
  {volume} {94}},\ \bibinfo {pages} {121113(R)} (\bibinfo {year}
  {2016})}\BibitemShut {NoStop}%
\bibitem [{\citenamefont {Bruno}\ \emph {et~al.}(2016)\citenamefont {Bruno},
  \citenamefont {Tamai}, \citenamefont {Wu}, \citenamefont {Cucchi},
  \citenamefont {Barreteau}, \citenamefont {de~la Torre}, \citenamefont
  {{McKeown Walker}}, \citenamefont {Ricc{\`{o}}}, \citenamefont {Wang},
  \citenamefont {Kim}, \citenamefont {Hoesch}, \citenamefont {Shi},
  \citenamefont {Plumb}, \citenamefont {Giannini}, \citenamefont {Soluyanov},\
  and\ \citenamefont {Baumberger}}]{Bruno:2016}%
  \BibitemOpen
  \bibfield  {author} {\bibinfo {author} {\bibfnamefont {F.~Y.}\ \bibnamefont
  {Bruno}}, \bibinfo {author} {\bibfnamefont {A.}~\bibnamefont {Tamai}},
  \bibinfo {author} {\bibfnamefont {Q.~S.}\ \bibnamefont {Wu}}, \bibinfo
  {author} {\bibfnamefont {I.}~\bibnamefont {Cucchi}}, \bibinfo {author}
  {\bibfnamefont {C.}~\bibnamefont {Barreteau}}, \bibinfo {author}
  {\bibfnamefont {A.}~\bibnamefont {de~la Torre}}, \bibinfo {author}
  {\bibfnamefont {S.}~\bibnamefont {{McKeown Walker}}}, \bibinfo {author}
  {\bibfnamefont {S.}~\bibnamefont {Ricc{\`{o}}}}, \bibinfo {author}
  {\bibfnamefont {Z.}~\bibnamefont {Wang}}, \bibinfo {author} {\bibfnamefont
  {T.~K.}\ \bibnamefont {Kim}}, \bibinfo {author} {\bibfnamefont
  {M.}~\bibnamefont {Hoesch}}, \bibinfo {author} {\bibfnamefont
  {M.}~\bibnamefont {Shi}}, \bibinfo {author} {\bibfnamefont {N.~C.}\
  \bibnamefont {Plumb}}, \bibinfo {author} {\bibfnamefont {E.}~\bibnamefont
  {Giannini}}, \bibinfo {author} {\bibfnamefont {A.~A.}\ \bibnamefont
  {Soluyanov}}, \ and\ \bibinfo {author} {\bibfnamefont {F.}~\bibnamefont
  {Baumberger}},\ }\href {\doibase 10.1103/PhysRevB.94.121112} {\bibfield
  {journal} {\bibinfo  {journal} {Phys. Rev. B}\ }\textbf {\bibinfo {volume}
  {94}},\ \bibinfo {pages} {121112(R)} (\bibinfo {year} {2016})}\BibitemShut
  {NoStop}%
\bibitem [{\citenamefont {{Di Sante}}\ \emph {et~al.}(2017)\citenamefont {{Di
  Sante}}, \citenamefont {Das}, \citenamefont {Bigi}, \citenamefont
  {Erg{\"{o}}nenc}, \citenamefont {G{\"{u}}rtler}, \citenamefont {Krieger},
  \citenamefont {Schmitt}, \citenamefont {Ali}, \citenamefont {Rossi},
  \citenamefont {Thomale}, \citenamefont {Franchini}, \citenamefont {Picozzi},
  \citenamefont {Fujii}, \citenamefont {Strocov}, \citenamefont {Sangiovanni},
  \citenamefont {Vobornik}, \citenamefont {Cava},\ and\ \citenamefont
  {Panaccione}}]{DiSante:2017}%
  \BibitemOpen
  \bibfield  {author} {\bibinfo {author} {\bibfnamefont {D.}~\bibnamefont {{Di
  Sante}}}, \bibinfo {author} {\bibfnamefont {P.~K.}\ \bibnamefont {Das}},
  \bibinfo {author} {\bibfnamefont {C.}~\bibnamefont {Bigi}}, \bibinfo {author}
  {\bibfnamefont {Z.}~\bibnamefont {Erg{\"{o}}nenc}}, \bibinfo {author}
  {\bibfnamefont {N.}~\bibnamefont {G{\"{u}}rtler}}, \bibinfo {author}
  {\bibfnamefont {J.~A.}\ \bibnamefont {Krieger}}, \bibinfo {author}
  {\bibfnamefont {T.}~\bibnamefont {Schmitt}}, \bibinfo {author} {\bibfnamefont
  {M.~N.}\ \bibnamefont {Ali}}, \bibinfo {author} {\bibfnamefont
  {G.}~\bibnamefont {Rossi}}, \bibinfo {author} {\bibfnamefont
  {R.}~\bibnamefont {Thomale}}, \bibinfo {author} {\bibfnamefont
  {C.}~\bibnamefont {Franchini}}, \bibinfo {author} {\bibfnamefont
  {S.}~\bibnamefont {Picozzi}}, \bibinfo {author} {\bibfnamefont
  {J.}~\bibnamefont {Fujii}}, \bibinfo {author} {\bibfnamefont {V.~N.}\
  \bibnamefont {Strocov}}, \bibinfo {author} {\bibfnamefont {G.}~\bibnamefont
  {Sangiovanni}}, \bibinfo {author} {\bibfnamefont {I.}~\bibnamefont
  {Vobornik}}, \bibinfo {author} {\bibfnamefont {R.~J.}\ \bibnamefont {Cava}},
  \ and\ \bibinfo {author} {\bibfnamefont {G.}~\bibnamefont {Panaccione}},\
  }\href {\doibase 10.1103/PhysRevLett.119.026403} {\bibfield  {journal}
  {\bibinfo  {journal} {Phys. Rev. Lett.}\ }\textbf {\bibinfo {volume} {119}},\
  \bibinfo {pages} {26403} (\bibinfo {year} {2017})}\BibitemShut {NoStop}%
\bibitem [{\citenamefont {Das}\ \emph {et~al.}(2019)\citenamefont {Das},
  \citenamefont {Sante}, \citenamefont {Cilento}, \citenamefont {Bigi},
  \citenamefont {Kopic}, \citenamefont {Soranzio}, \citenamefont {Sterzi},
  \citenamefont {Krieger}, \citenamefont {Vobornik}, \citenamefont {Fujii},
  \citenamefont {Okuda}, \citenamefont {Strocov}, \citenamefont {Breese},
  \citenamefont {Parmigiani}, \citenamefont {Rossi}, \citenamefont {Picozzi},
  \citenamefont {Thomale}, \citenamefont {Sangiovanni}, \citenamefont {Cava},\
  and\ \citenamefont {Panaccione}}]{Das:2019}%
  \BibitemOpen
  \bibfield  {author} {\bibinfo {author} {\bibfnamefont {P.~K.}\ \bibnamefont
  {Das}}, \bibinfo {author} {\bibfnamefont {D.~D.}\ \bibnamefont {Sante}},
  \bibinfo {author} {\bibfnamefont {F.}~\bibnamefont {Cilento}}, \bibinfo
  {author} {\bibfnamefont {C.}~\bibnamefont {Bigi}}, \bibinfo {author}
  {\bibfnamefont {D.}~\bibnamefont {Kopic}}, \bibinfo {author} {\bibfnamefont
  {D.}~\bibnamefont {Soranzio}}, \bibinfo {author} {\bibfnamefont
  {A.}~\bibnamefont {Sterzi}}, \bibinfo {author} {\bibfnamefont {J.~A.}\
  \bibnamefont {Krieger}}, \bibinfo {author} {\bibfnamefont {I.}~\bibnamefont
  {Vobornik}}, \bibinfo {author} {\bibfnamefont {J.}~\bibnamefont {Fujii}},
  \bibinfo {author} {\bibfnamefont {T.}~\bibnamefont {Okuda}}, \bibinfo
  {author} {\bibfnamefont {V.~N.}\ \bibnamefont {Strocov}}, \bibinfo {author}
  {\bibfnamefont {M.~B.~H.}\ \bibnamefont {Breese}}, \bibinfo {author}
  {\bibfnamefont {F.}~\bibnamefont {Parmigiani}}, \bibinfo {author}
  {\bibfnamefont {G.}~\bibnamefont {Rossi}}, \bibinfo {author} {\bibfnamefont
  {S.}~\bibnamefont {Picozzi}}, \bibinfo {author} {\bibfnamefont
  {R.}~\bibnamefont {Thomale}}, \bibinfo {author} {\bibfnamefont
  {G.}~\bibnamefont {Sangiovanni}}, \bibinfo {author} {\bibfnamefont {R.~J.}\
  \bibnamefont {Cava}}, \ and\ \bibinfo {author} {\bibfnamefont
  {G.}~\bibnamefont {Panaccione}},\ }\href {\doibase 10.1088/2516-1075/ab0835}
  {\bibfield  {journal} {\bibinfo  {journal} {Electronic Structure}\ }\textbf
  {\bibinfo {volume} {1}},\ \bibinfo {pages} {14003} (\bibinfo {year}
  {2019})}\BibitemShut {NoStop}%
\bibitem [{\citenamefont {Kononov}\ \emph {et~al.}(2018)\citenamefont
  {Kononov}, \citenamefont {Shvetsov}, \citenamefont {Egorov}, \citenamefont
  {Timonina}, \citenamefont {Kolesnikov},\ and\ \citenamefont
  {Deviatov}}]{Kononov:2018}%
  \BibitemOpen
  \bibfield  {author} {\bibinfo {author} {\bibfnamefont {A.}~\bibnamefont
  {Kononov}}, \bibinfo {author} {\bibfnamefont {O.~O.}\ \bibnamefont
  {Shvetsov}}, \bibinfo {author} {\bibfnamefont {S.~V.}\ \bibnamefont
  {Egorov}}, \bibinfo {author} {\bibfnamefont {A.~V.}\ \bibnamefont
  {Timonina}}, \bibinfo {author} {\bibfnamefont {N.~N.}\ \bibnamefont
  {Kolesnikov}}, \ and\ \bibinfo {author} {\bibfnamefont {E.~V.}\ \bibnamefont
  {Deviatov}},\ }\href {\doibase 10.1209/0295-5075/122/27004} {\bibfield
  {journal} {\bibinfo  {journal} {EPL}\ }\textbf {\bibinfo {volume} {122}},\
  \bibinfo {pages} {27004} (\bibinfo {year} {2018})}\BibitemShut {NoStop}%
\bibitem [{\citenamefont {Li}\ \emph {et~al.}(2018)\citenamefont {Li},
  \citenamefont {Wu}, \citenamefont {Wen}, \citenamefont {Zhang}, \citenamefont
  {Zhang}, \citenamefont {Zhang}, \citenamefont {Yu}, \citenamefont {Yang},
  \citenamefont {Manchon},\ and\ \citenamefont {Zhang}}]{Li:2018}%
  \BibitemOpen
  \bibfield  {author} {\bibinfo {author} {\bibfnamefont {P.}~\bibnamefont
  {Li}}, \bibinfo {author} {\bibfnamefont {W.}~\bibnamefont {Wu}}, \bibinfo
  {author} {\bibfnamefont {Y.}~\bibnamefont {Wen}}, \bibinfo {author}
  {\bibfnamefont {C.}~\bibnamefont {Zhang}}, \bibinfo {author} {\bibfnamefont
  {J.}~\bibnamefont {Zhang}}, \bibinfo {author} {\bibfnamefont
  {S.}~\bibnamefont {Zhang}}, \bibinfo {author} {\bibfnamefont
  {Z.}~\bibnamefont {Yu}}, \bibinfo {author} {\bibfnamefont {S.~A.}\
  \bibnamefont {Yang}}, \bibinfo {author} {\bibfnamefont {A.}~\bibnamefont
  {Manchon}}, \ and\ \bibinfo {author} {\bibfnamefont {X.-x.}\ \bibnamefont
  {Zhang}},\ }\href {\doibase 10.1038/s41467-018-06518-1} {\bibfield  {journal}
  {\bibinfo  {journal} {Nature Communications}\ }\textbf {\bibinfo {volume}
  {9}},\ \bibinfo {pages} {3990} (\bibinfo {year} {2018})}\BibitemShut
  {NoStop}%
\bibitem [{\citenamefont {Feng}\ \emph {et~al.}(2016)\citenamefont {Feng},
  \citenamefont {Chan}, \citenamefont {Feng}, \citenamefont {Liu},
  \citenamefont {Chou}, \citenamefont {Kuroda}, \citenamefont {Yaji},
  \citenamefont {Harasawa}, \citenamefont {Moras}, \citenamefont {Barinov},
  \citenamefont {Malaeb}, \citenamefont {Bareille}, \citenamefont {Kondo},
  \citenamefont {Shin}, \citenamefont {Komori}, \citenamefont {Chiang},
  \citenamefont {Shi},\ and\ \citenamefont {Matsuda}}]{Feng:2016}%
  \BibitemOpen
  \bibfield  {author} {\bibinfo {author} {\bibfnamefont {B.}~\bibnamefont
  {Feng}}, \bibinfo {author} {\bibfnamefont {Y.-H.}\ \bibnamefont {Chan}},
  \bibinfo {author} {\bibfnamefont {Y.}~\bibnamefont {Feng}}, \bibinfo {author}
  {\bibfnamefont {R.-Y.}\ \bibnamefont {Liu}}, \bibinfo {author} {\bibfnamefont
  {M.-Y.}\ \bibnamefont {Chou}}, \bibinfo {author} {\bibfnamefont
  {K.}~\bibnamefont {Kuroda}}, \bibinfo {author} {\bibfnamefont
  {K.}~\bibnamefont {Yaji}}, \bibinfo {author} {\bibfnamefont {A.}~\bibnamefont
  {Harasawa}}, \bibinfo {author} {\bibfnamefont {P.}~\bibnamefont {Moras}},
  \bibinfo {author} {\bibfnamefont {A.}~\bibnamefont {Barinov}}, \bibinfo
  {author} {\bibfnamefont {W.}~\bibnamefont {Malaeb}}, \bibinfo {author}
  {\bibfnamefont {C.}~\bibnamefont {Bareille}}, \bibinfo {author}
  {\bibfnamefont {T.}~\bibnamefont {Kondo}}, \bibinfo {author} {\bibfnamefont
  {S.}~\bibnamefont {Shin}}, \bibinfo {author} {\bibfnamefont {F.}~\bibnamefont
  {Komori}}, \bibinfo {author} {\bibfnamefont {T.-C.}\ \bibnamefont {Chiang}},
  \bibinfo {author} {\bibfnamefont {Y.}~\bibnamefont {Shi}}, \ and\ \bibinfo
  {author} {\bibfnamefont {I.}~\bibnamefont {Matsuda}},\ }\href {\doibase
  10.1103/PhysRevB.94.195134} {\bibfield  {journal} {\bibinfo  {journal} {Phys.
  Rev. B}\ }\textbf {\bibinfo {volume} {94}},\ \bibinfo {pages} {195134}
  (\bibinfo {year} {2016})}\BibitemShut {NoStop}%
\bibitem [{\citenamefont {Caputo}\ \emph {et~al.}(2018)\citenamefont {Caputo},
  \citenamefont {Khalil}, \citenamefont {Papalazarou}, \citenamefont
  {Nilforoushan}, \citenamefont {Perfetti}, \citenamefont {Taleb-Ibrahimi},
  \citenamefont {Gibson}, \citenamefont {Cava},\ and\ \citenamefont
  {Marsi}}]{Caputo:2018}%
  \BibitemOpen
  \bibfield  {author} {\bibinfo {author} {\bibfnamefont {M.}~\bibnamefont
  {Caputo}}, \bibinfo {author} {\bibfnamefont {L.}~\bibnamefont {Khalil}},
  \bibinfo {author} {\bibfnamefont {E.}~\bibnamefont {Papalazarou}}, \bibinfo
  {author} {\bibfnamefont {N.}~\bibnamefont {Nilforoushan}}, \bibinfo {author}
  {\bibfnamefont {L.}~\bibnamefont {Perfetti}}, \bibinfo {author}
  {\bibfnamefont {A.}~\bibnamefont {Taleb-Ibrahimi}}, \bibinfo {author}
  {\bibfnamefont {Q.~D.}\ \bibnamefont {Gibson}}, \bibinfo {author}
  {\bibfnamefont {R.~J.}\ \bibnamefont {Cava}}, \ and\ \bibinfo {author}
  {\bibfnamefont {M.}~\bibnamefont {Marsi}},\ }\href {\doibase
  10.1103/PhysRevB.97.115115} {\bibfield  {journal} {\bibinfo  {journal} {Phys.
  Rev. B}\ }\textbf {\bibinfo {volume} {97}},\ \bibinfo {pages} {115115}
  (\bibinfo {year} {2018})}\BibitemShut {NoStop}%
\bibitem [{\citenamefont {Scholl}\ \emph {et~al.}(1997)\citenamefont {Scholl},
  \citenamefont {Baumgarten}, \citenamefont {Jacquemin},\ and\ \citenamefont
  {Eberhardt}}]{Scholl:1997}%
  \BibitemOpen
  \bibfield  {author} {\bibinfo {author} {\bibfnamefont {A.}~\bibnamefont
  {Scholl}}, \bibinfo {author} {\bibfnamefont {L.}~\bibnamefont {Baumgarten}},
  \bibinfo {author} {\bibfnamefont {R.}~\bibnamefont {Jacquemin}}, \ and\
  \bibinfo {author} {\bibfnamefont {W.}~\bibnamefont {Eberhardt}},\ }\href
  {\doibase 10.1103/PhysRevLett.79.5146} {\bibfield  {journal} {\bibinfo
  {journal} {Phys. Rev. Lett.}\ }\textbf {\bibinfo {volume} {79}},\ \bibinfo
  {pages} {5146} (\bibinfo {year} {1997})}\BibitemShut {NoStop}%
\bibitem [{\citenamefont {Cinchetti}\ \emph {et~al.}(2006)\citenamefont
  {Cinchetti}, \citenamefont {{S{\'{a}}nchez Albaneda}}, \citenamefont
  {Hoffmann}, \citenamefont {Roth}, \citenamefont {W{\"{u}}stenberg},
  \citenamefont {Krau{\ss}}, \citenamefont {Andreyev}, \citenamefont {Schneider},
  \citenamefont {Bauer},\ and\ \citenamefont {Aeschlimann}}]{Cinchetti:2006}%
  \BibitemOpen
  \bibfield  {author} {\bibinfo {author} {\bibfnamefont {M.}~\bibnamefont
  {Cinchetti}}, \bibinfo {author} {\bibfnamefont {M.}~\bibnamefont
  {{S{\'{a}}nchez Albaneda}}}, \bibinfo {author} {\bibfnamefont
  {D.}~\bibnamefont {Hoffmann}}, \bibinfo {author} {\bibfnamefont
  {T.}~\bibnamefont {Roth}}, \bibinfo {author} {\bibfnamefont {J.-P.}\
  \bibnamefont {W{\"{u}}stenberg}}, \bibinfo {author} {\bibfnamefont
  {M.}~\bibnamefont {Krauss}}, \bibinfo {author} {\bibfnamefont
  {O.}~\bibnamefont {Andreyev}}, \bibinfo {author} {\bibfnamefont {H.~C.}\
  \bibnamefont {Schneider}}, \bibinfo {author} {\bibfnamefont {M.}~\bibnamefont
  {Bauer}}, \ and\ \bibinfo {author} {\bibfnamefont {M.}~\bibnamefont
  {Aeschlimann}},\ }\href {\doibase 10.1103/PhysRevLett.97.177201} {\bibfield
  {journal} {\bibinfo  {journal} {Phys. Rev. Lett.}\ }\textbf {\bibinfo
  {volume} {97}},\ \bibinfo {pages} {177201} (\bibinfo {year}
  {2006})}\BibitemShut {NoStop}%
\bibitem [{\citenamefont {Weber}\ \emph {et~al.}(2011)\citenamefont {Weber},
  \citenamefont {Pressacco}, \citenamefont {G{\"{u}}nther}, \citenamefont
  {Mancini}, \citenamefont {Oppeneer},\ and\ \citenamefont
  {Back}}]{Weber:2011}%
  \BibitemOpen
  \bibfield  {author} {\bibinfo {author} {\bibfnamefont {A.}~\bibnamefont
  {Weber}}, \bibinfo {author} {\bibfnamefont {F.}~\bibnamefont {Pressacco}},
  \bibinfo {author} {\bibfnamefont {S.}~\bibnamefont {G{\"{u}}nther}}, \bibinfo
  {author} {\bibfnamefont {E.}~\bibnamefont {Mancini}}, \bibinfo {author}
  {\bibfnamefont {P.~M.}\ \bibnamefont {Oppeneer}}, \ and\ \bibinfo {author}
  {\bibfnamefont {C.~H.}\ \bibnamefont {Back}},\ }\href {\doibase
  10.1103/PhysRevB.84.132412} {\bibfield  {journal} {\bibinfo  {journal} {Phys.
  Rev. B}\ }\textbf {\bibinfo {volume} {84}},\ \bibinfo {pages} {132412}
  (\bibinfo {year} {2011})}\BibitemShut {NoStop}%
\bibitem [{\citenamefont {Cacho}\ \emph {et~al.}(2015)\citenamefont {Cacho},
  \citenamefont {Crepaldi}, \citenamefont {Battiato}, \citenamefont {Braun},
  \citenamefont {Cilento}, \citenamefont {Zacchigna}, \citenamefont {Richter},
  \citenamefont {Heckmann}, \citenamefont {Springate}, \citenamefont {Liu},
  \citenamefont {Dhesi}, \citenamefont {Berger}, \citenamefont {Bugnon},
  \citenamefont {Held}, \citenamefont {Grioni}, \citenamefont {Ebert},
  \citenamefont {Hricovini}, \citenamefont {Min{\'{a}}r},\ and\ \citenamefont
  {Parmigiani}}]{Cacho:2015}%
  \BibitemOpen
  \bibfield  {author} {\bibinfo {author} {\bibfnamefont {C.}~\bibnamefont
  {Cacho}}, \bibinfo {author} {\bibfnamefont {A.}~\bibnamefont {Crepaldi}},
  \bibinfo {author} {\bibfnamefont {M.}~\bibnamefont {Battiato}}, \bibinfo
  {author} {\bibfnamefont {J.}~\bibnamefont {Braun}}, \bibinfo {author}
  {\bibfnamefont {F.}~\bibnamefont {Cilento}}, \bibinfo {author} {\bibfnamefont
  {M.}~\bibnamefont {Zacchigna}}, \bibinfo {author} {\bibfnamefont {M.~C.}\
  \bibnamefont {Richter}}, \bibinfo {author} {\bibfnamefont {O.}~\bibnamefont
  {Heckmann}}, \bibinfo {author} {\bibfnamefont {E.}~\bibnamefont {Springate}},
  \bibinfo {author} {\bibfnamefont {Y.}~\bibnamefont {Liu}}, \bibinfo {author}
  {\bibfnamefont {S.~S.}\ \bibnamefont {Dhesi}}, \bibinfo {author}
  {\bibfnamefont {H.}~\bibnamefont {Berger}}, \bibinfo {author} {\bibfnamefont
  {P.}~\bibnamefont {Bugnon}}, \bibinfo {author} {\bibfnamefont
  {K.}~\bibnamefont {Held}}, \bibinfo {author} {\bibfnamefont {M.}~\bibnamefont
  {Grioni}}, \bibinfo {author} {\bibfnamefont {H.}~\bibnamefont {Ebert}},
  \bibinfo {author} {\bibfnamefont {K.}~\bibnamefont {Hricovini}}, \bibinfo
  {author} {\bibfnamefont {J.}~\bibnamefont {Min{\'{a}}r}}, \ and\ \bibinfo
  {author} {\bibfnamefont {F.}~\bibnamefont {Parmigiani}},\ }\href {\doibase
  10.1103/PhysRevLett.114.097401} {\bibfield  {journal} {\bibinfo  {journal}
  {Phys. Rev. Lett.}\ }\textbf {\bibinfo {volume} {114}},\ \bibinfo {pages}
  {97401} (\bibinfo {year} {2015})}\BibitemShut {NoStop}%
\bibitem [{\citenamefont {S{\'{a}}nchez-Barriga}\ \emph
  {et~al.}(2016)\citenamefont {S{\'{a}}nchez-Barriga}, \citenamefont {Golias},
  \citenamefont {Varykhalov}, \citenamefont {Braun}, \citenamefont {Yashina},
  \citenamefont {Schumann}, \citenamefont {Min{\'{a}}r}, \citenamefont {Ebert},
  \citenamefont {Kornilov},\ and\ \citenamefont {Rader}}]{Barriga:2016}%
  \BibitemOpen
  \bibfield  {author} {\bibinfo {author} {\bibfnamefont {J.}~\bibnamefont
  {S{\'{a}}nchez-Barriga}}, \bibinfo {author} {\bibfnamefont {E.}~\bibnamefont
  {Golias}}, \bibinfo {author} {\bibfnamefont {A.}~\bibnamefont {Varykhalov}},
  \bibinfo {author} {\bibfnamefont {J.}~\bibnamefont {Braun}}, \bibinfo
  {author} {\bibfnamefont {L.~V.}\ \bibnamefont {Yashina}}, \bibinfo {author}
  {\bibfnamefont {R.}~\bibnamefont {Schumann}}, \bibinfo {author}
  {\bibfnamefont {J.}~\bibnamefont {Min{\'{a}}r}}, \bibinfo {author}
  {\bibfnamefont {H.}~\bibnamefont {Ebert}}, \bibinfo {author} {\bibfnamefont
  {O.}~\bibnamefont {Kornilov}}, \ and\ \bibinfo {author} {\bibfnamefont
  {O.}~\bibnamefont {Rader}},\ }\href {\doibase 10.1103/PhysRevB.93.155426}
  {\bibfield  {journal} {\bibinfo  {journal} {Phys. Rev. B}\ }\textbf {\bibinfo
  {volume} {93}},\ \bibinfo {pages} {155426} (\bibinfo {year}
  {2016})}\BibitemShut {NoStop}%
\bibitem [{\citenamefont {Battiato}\ \emph {et~al.}(2018)\citenamefont
  {Battiato}, \citenamefont {Min{\'{a}}r}, \citenamefont {Wang}, \citenamefont
  {Ndiaye}, \citenamefont {Richter}, \citenamefont {Heckmann}, \citenamefont
  {Mariot}, \citenamefont {Parmigiani}, \citenamefont {Hricovini},\ and\
  \citenamefont {Cacho}}]{Battiato:2018}%
  \BibitemOpen
  \bibfield  {author} {\bibinfo {author} {\bibfnamefont {M.}~\bibnamefont
  {Battiato}}, \bibinfo {author} {\bibfnamefont {J.}~\bibnamefont
  {Min{\'{a}}r}}, \bibinfo {author} {\bibfnamefont {W.}~\bibnamefont {Wang}},
  \bibinfo {author} {\bibfnamefont {W.}~\bibnamefont {Ndiaye}}, \bibinfo
  {author} {\bibfnamefont {M.~C.}\ \bibnamefont {Richter}}, \bibinfo {author}
  {\bibfnamefont {O.}~\bibnamefont {Heckmann}}, \bibinfo {author}
  {\bibfnamefont {J.-M.}\ \bibnamefont {Mariot}}, \bibinfo {author}
  {\bibfnamefont {F.}~\bibnamefont {Parmigiani}}, \bibinfo {author}
  {\bibfnamefont {K.}~\bibnamefont {Hricovini}}, \ and\ \bibinfo {author}
  {\bibfnamefont {C.}~\bibnamefont {Cacho}},\ }\href {\doibase
  10.1103/PhysRevLett.121.077205} {\bibfield  {journal} {\bibinfo  {journal}
  {Phys. Rev. Lett.}\ }\textbf {\bibinfo {volume} {121}},\ \bibinfo {pages}
  {77205} (\bibinfo {year} {2018})}\BibitemShut {NoStop}%
\bibitem [{\citenamefont {Fognini}\ \emph {et~al.}(2014)\citenamefont
  {Fognini}, \citenamefont {Michlmayr}, \citenamefont {Salvatella},
  \citenamefont {Wetli}, \citenamefont {Ramsperger}, \citenamefont
  {B{\"{a}}hler}, \citenamefont {Sorgenfrei}, \citenamefont {Beye},
  \citenamefont {Eschenlohr}, \citenamefont {Pontius}, \citenamefont {Stamm},
  \citenamefont {Hieke}, \citenamefont {Dell'Angela}, \citenamefont {de~Jong},
  \citenamefont {Kukreja}, \citenamefont {Gerasimova}, \citenamefont
  {Rybnikov}, \citenamefont {Al-Shemmary}, \citenamefont {Redlin},
  \citenamefont {Raabe}, \citenamefont {F{\"{o}}hlisch}, \citenamefont
  {D{\"{u}}rr}, \citenamefont {Wurth}, \citenamefont {Pescia}, \citenamefont
  {Vaterlaus},\ and\ \citenamefont {Acremann}}]{Fognini:2014}%
  \BibitemOpen
  \bibfield  {author} {\bibinfo {author} {\bibfnamefont {A.}~\bibnamefont
  {Fognini}}, \bibinfo {author} {\bibfnamefont {T.~U.}\ \bibnamefont
  {Michlmayr}}, \bibinfo {author} {\bibfnamefont {G.}~\bibnamefont
  {Salvatella}}, \bibinfo {author} {\bibfnamefont {C.}~\bibnamefont {Wetli}},
  \bibinfo {author} {\bibfnamefont {U.}~\bibnamefont {Ramsperger}}, \bibinfo
  {author} {\bibfnamefont {T.}~\bibnamefont {B{\"{a}}hler}}, \bibinfo {author}
  {\bibfnamefont {F.}~\bibnamefont {Sorgenfrei}}, \bibinfo {author}
  {\bibfnamefont {M.}~\bibnamefont {Beye}}, \bibinfo {author} {\bibfnamefont
  {A.}~\bibnamefont {Eschenlohr}}, \bibinfo {author} {\bibfnamefont
  {N.}~\bibnamefont {Pontius}}, \bibinfo {author} {\bibfnamefont
  {C.}~\bibnamefont {Stamm}}, \bibinfo {author} {\bibfnamefont
  {F.}~\bibnamefont {Hieke}}, \bibinfo {author} {\bibfnamefont
  {M.}~\bibnamefont {Dell'Angela}}, \bibinfo {author} {\bibfnamefont
  {S.}~\bibnamefont {de~Jong}}, \bibinfo {author} {\bibfnamefont
  {R.}~\bibnamefont {Kukreja}}, \bibinfo {author} {\bibfnamefont
  {N.}~\bibnamefont {Gerasimova}}, \bibinfo {author} {\bibfnamefont
  {V.}~\bibnamefont {Rybnikov}}, \bibinfo {author} {\bibfnamefont
  {A.}~\bibnamefont {Al-Shemmary}}, \bibinfo {author} {\bibfnamefont
  {H.}~\bibnamefont {Redlin}}, \bibinfo {author} {\bibfnamefont
  {J.}~\bibnamefont {Raabe}}, \bibinfo {author} {\bibfnamefont
  {A.}~\bibnamefont {F{\"{o}}hlisch}}, \bibinfo {author} {\bibfnamefont
  {H.~A.}\ \bibnamefont {D{\"{u}}rr}}, \bibinfo {author} {\bibfnamefont
  {W.}~\bibnamefont {Wurth}}, \bibinfo {author} {\bibfnamefont
  {D.}~\bibnamefont {Pescia}}, \bibinfo {author} {\bibfnamefont
  {A.}~\bibnamefont {Vaterlaus}}, \ and\ \bibinfo {author} {\bibfnamefont
  {Y.}~\bibnamefont {Acremann}},\ }\href {\doibase 10.1063/1.4862476}
  {\bibfield  {journal} {\bibinfo  {journal} {Applied Physics Letters}\
  }\textbf {\bibinfo {volume} {104}},\ \bibinfo {pages} {32402} (\bibinfo
  {year} {2014})}\BibitemShut {NoStop}%
\bibitem [{\citenamefont {Pl{\"{o}}tzing}\ \emph {et~al.}(2016)\citenamefont
  {Pl{\"{o}}tzing}, \citenamefont {Adam}, \citenamefont {Weier}, \citenamefont
  {Plucinski}, \citenamefont {Eich}, \citenamefont {Emmerich}, \citenamefont
  {Rollinger}, \citenamefont {Aeschlimann}, \citenamefont {Mathias},\ and\
  \citenamefont {Schneider}}]{Plotzing:2016}%
  \BibitemOpen
  \bibfield  {author} {\bibinfo {author} {\bibfnamefont {M.}~\bibnamefont
  {Pl{\"{o}}tzing}}, \bibinfo {author} {\bibfnamefont {R.}~\bibnamefont
  {Adam}}, \bibinfo {author} {\bibfnamefont {C.}~\bibnamefont {Weier}},
  \bibinfo {author} {\bibfnamefont {L.}~\bibnamefont {Plucinski}}, \bibinfo
  {author} {\bibfnamefont {S.}~\bibnamefont {Eich}}, \bibinfo {author}
  {\bibfnamefont {S.}~\bibnamefont {Emmerich}}, \bibinfo {author}
  {\bibfnamefont {M.}~\bibnamefont {Rollinger}}, \bibinfo {author}
  {\bibfnamefont {M.}~\bibnamefont {Aeschlimann}}, \bibinfo {author}
  {\bibfnamefont {S.}~\bibnamefont {Mathias}}, \ and\ \bibinfo {author}
  {\bibfnamefont {C.~M.}\ \bibnamefont {Schneider}},\ }\href {\doibase
  10.1063/1.4946782} {\bibfield  {journal} {\bibinfo  {journal} {Review of
  Scientific Instruments}\ }\textbf {\bibinfo {volume} {87}},\ \bibinfo {pages}
  {43903} (\bibinfo {year} {2016})}\BibitemShut {NoStop}%
\bibitem [{\citenamefont {Eich}\ \emph {et~al.}(2017)\citenamefont {Eich},
  \citenamefont {Pl{\"{o}}tzing}, \citenamefont {Rollinger}, \citenamefont
  {Emmerich}, \citenamefont {Adam}, \citenamefont {Chen}, \citenamefont
  {Kapteyn}, \citenamefont {Murnane}, \citenamefont {Plucinski}, \citenamefont
  {Steil}, \citenamefont {Stadtm{\"{u}}ller}, \citenamefont {Cinchetti},
  \citenamefont {Aeschlimann}, \citenamefont {Schneider},\ and\ \citenamefont
  {Mathias}}]{Eich:2017}%
  \BibitemOpen
  \bibfield  {author} {\bibinfo {author} {\bibfnamefont {S.}~\bibnamefont
  {Eich}}, \bibinfo {author} {\bibfnamefont {M.}~\bibnamefont
  {Pl{\"{o}}tzing}}, \bibinfo {author} {\bibfnamefont {M.}~\bibnamefont
  {Rollinger}}, \bibinfo {author} {\bibfnamefont {S.}~\bibnamefont {Emmerich}},
  \bibinfo {author} {\bibfnamefont {R.}~\bibnamefont {Adam}}, \bibinfo {author}
  {\bibfnamefont {C.}~\bibnamefont {Chen}}, \bibinfo {author} {\bibfnamefont
  {H.~C.}\ \bibnamefont {Kapteyn}}, \bibinfo {author} {\bibfnamefont {M.~M.}\
  \bibnamefont {Murnane}}, \bibinfo {author} {\bibfnamefont {L.}~\bibnamefont
  {Plucinski}}, \bibinfo {author} {\bibfnamefont {D.}~\bibnamefont {Steil}},
  \bibinfo {author} {\bibfnamefont {B.}~\bibnamefont {Stadtm{\"{u}}ller}},
  \bibinfo {author} {\bibfnamefont {M.}~\bibnamefont {Cinchetti}}, \bibinfo
  {author} {\bibfnamefont {M.}~\bibnamefont {Aeschlimann}}, \bibinfo {author}
  {\bibfnamefont {C.~M.}\ \bibnamefont {Schneider}}, \ and\ \bibinfo {author}
  {\bibfnamefont {S.}~\bibnamefont {Mathias}},\ }\href
  {https://advances.sciencemag.org/content/3/3/e1602094} {\bibfield  {journal}
  {\bibinfo  {journal} {Science Advances}\ }\textbf {\bibinfo {volume} {3}}
  (\bibinfo {year} {2017})}\BibitemShut {NoStop}%
\bibitem [{\citenamefont {Nie}\ \emph {et~al.}(2019)\citenamefont {Nie},
  \citenamefont {Turcu}, \citenamefont {Li}, \citenamefont {Zhang},
  \citenamefont {He}, \citenamefont {Tu}, \citenamefont {Ni}, \citenamefont
  {Xu}, \citenamefont {Chen}, \citenamefont {Ruan}, \citenamefont {Frassetto},
  \citenamefont {Miotti}, \citenamefont {Fabris}, \citenamefont {Poletto},
  \citenamefont {Wu}, \citenamefont {Lu}, \citenamefont {Liu}, \citenamefont
  {Kampen}, \citenamefont {Zhai}, \citenamefont {Liu}, \citenamefont {Cacho},
  \citenamefont {Wang}, \citenamefont {Wang}, \citenamefont {Shi},
  \citenamefont {Zhang},\ and\ \citenamefont {Xu}}]{Nie:2019}%
  \BibitemOpen
  \bibfield  {author} {\bibinfo {author} {\bibfnamefont {Z.}~\bibnamefont
  {Nie}}, \bibinfo {author} {\bibfnamefont {I.~C.~E.}\ \bibnamefont {Turcu}},
  \bibinfo {author} {\bibfnamefont {Y.}~\bibnamefont {Li}}, \bibinfo {author}
  {\bibfnamefont {X.}~\bibnamefont {Zhang}}, \bibinfo {author} {\bibfnamefont
  {L.}~\bibnamefont {He}}, \bibinfo {author} {\bibfnamefont {J.}~\bibnamefont
  {Tu}}, \bibinfo {author} {\bibfnamefont {Z.}~\bibnamefont {Ni}}, \bibinfo
  {author} {\bibfnamefont {H.}~\bibnamefont {Xu}}, \bibinfo {author}
  {\bibfnamefont {Y.}~\bibnamefont {Chen}}, \bibinfo {author} {\bibfnamefont
  {X.}~\bibnamefont {Ruan}}, \bibinfo {author} {\bibfnamefont {F.}~\bibnamefont
  {Frassetto}}, \bibinfo {author} {\bibfnamefont {P.}~\bibnamefont {Miotti}},
  \bibinfo {author} {\bibfnamefont {N.}~\bibnamefont {Fabris}}, \bibinfo
  {author} {\bibfnamefont {L.}~\bibnamefont {Poletto}}, \bibinfo {author}
  {\bibfnamefont {J.}~\bibnamefont {Wu}}, \bibinfo {author} {\bibfnamefont
  {Q.}~\bibnamefont {Lu}}, \bibinfo {author} {\bibfnamefont {C.}~\bibnamefont
  {Liu}}, \bibinfo {author} {\bibfnamefont {T.}~\bibnamefont {Kampen}},
  \bibinfo {author} {\bibfnamefont {Y.}~\bibnamefont {Zhai}}, \bibinfo {author}
  {\bibfnamefont {W.}~\bibnamefont {Liu}}, \bibinfo {author} {\bibfnamefont
  {C.}~\bibnamefont {Cacho}}, \bibinfo {author} {\bibfnamefont
  {X.}~\bibnamefont {Wang}}, \bibinfo {author} {\bibfnamefont {F.}~\bibnamefont
  {Wang}}, \bibinfo {author} {\bibfnamefont {Y.}~\bibnamefont {Shi}}, \bibinfo
  {author} {\bibfnamefont {R.}~\bibnamefont {Zhang}}, \ and\ \bibinfo {author}
  {\bibfnamefont {Y.}~\bibnamefont {Xu}},\ }\href {\doibase 10.3390/app9030370}
  {\bibfield  {journal} {\bibinfo  {journal} {Applied Sciences}\ }\textbf
  {\bibinfo {volume} {9}},\ \bibinfo {pages} {370} (\bibinfo {year}
  {2019})}\BibitemShut {NoStop}%
\bibitem [{\citenamefont {Golinelli}\ \emph {et~al.}(2017)\citenamefont
  {Golinelli}, \citenamefont {Chen}, \citenamefont {Gontier}, \citenamefont
  {Bussi{\`{e}}re}, \citenamefont {Tcherbakoff}, \citenamefont {Natile},
  \citenamefont {D'Oliveira}, \citenamefont {Paul},\ and\ \citenamefont
  {Hergott}}]{Golinelli:2017}%
  \BibitemOpen
  \bibfield  {author} {\bibinfo {author} {\bibfnamefont {A.}~\bibnamefont
  {Golinelli}}, \bibinfo {author} {\bibfnamefont {X.}~\bibnamefont {Chen}},
  \bibinfo {author} {\bibfnamefont {E.}~\bibnamefont {Gontier}}, \bibinfo
  {author} {\bibfnamefont {B.}~\bibnamefont {Bussi{\`{e}}re}}, \bibinfo
  {author} {\bibfnamefont {O.}~\bibnamefont {Tcherbakoff}}, \bibinfo {author}
  {\bibfnamefont {M.}~\bibnamefont {Natile}}, \bibinfo {author} {\bibfnamefont
  {P.}~\bibnamefont {D'Oliveira}}, \bibinfo {author} {\bibfnamefont {P.-M.}\
  \bibnamefont {Paul}}, \ and\ \bibinfo {author} {\bibfnamefont {J.-F.}\
  \bibnamefont {Hergott}},\ }\href {\doibase 10.1364/OL.42.002326} {\bibfield
  {journal} {\bibinfo  {journal} {Opt. Lett.}\ }\textbf {\bibinfo {volume}
  {42}},\ \bibinfo {pages} {2326} (\bibinfo {year} {2017})}\BibitemShut
  {NoStop}%
\bibitem [{\citenamefont {Golinelli}\ \emph {et~al.}(2019)\citenamefont
  {Golinelli}, \citenamefont {Chen}, \citenamefont {Bussi{\`{e}}re},
  \citenamefont {Gontier}, \citenamefont {Paul}, \citenamefont {Tcherbakoff},
  \citenamefont {D'Oliveira},\ and\ \citenamefont {Hergott}}]{Golinelli:2019}%
  \BibitemOpen
  \bibfield  {author} {\bibinfo {author} {\bibfnamefont {A.}~\bibnamefont
  {Golinelli}}, \bibinfo {author} {\bibfnamefont {X.}~\bibnamefont {Chen}},
  \bibinfo {author} {\bibfnamefont {B.}~\bibnamefont {Bussi{\`{e}}re}},
  \bibinfo {author} {\bibfnamefont {E.}~\bibnamefont {Gontier}}, \bibinfo
  {author} {\bibfnamefont {P.-M.}\ \bibnamefont {Paul}}, \bibinfo {author}
  {\bibfnamefont {O.}~\bibnamefont {Tcherbakoff}}, \bibinfo {author}
  {\bibfnamefont {P.}~\bibnamefont {D'Oliveira}}, \ and\ \bibinfo {author}
  {\bibfnamefont {J.-F.}\ \bibnamefont {Hergott}},\ }\href {\doibase
  10.1364/OE.27.013624} {\bibfield  {journal} {\bibinfo  {journal} {Opt.
  Express}\ }\textbf {\bibinfo {volume} {27}},\ \bibinfo {pages} {13624}
  (\bibinfo {year} {2019})}\BibitemShut {NoStop}%
\bibitem [{\citenamefont {Frassetto}\ \emph {et~al.}(2011)\citenamefont
  {Frassetto}, \citenamefont {Cacho}, \citenamefont {Froud}, \citenamefont
  {Turcu}, \citenamefont {Villoresi}, \citenamefont {Bryan}, \citenamefont
  {Springate},\ and\ \citenamefont {Poletto}}]{Frassetto:2011}%
  \BibitemOpen
  \bibfield  {author} {\bibinfo {author} {\bibfnamefont {F.}~\bibnamefont
  {Frassetto}}, \bibinfo {author} {\bibfnamefont {C.}~\bibnamefont {Cacho}},
  \bibinfo {author} {\bibfnamefont {C.~A.}\ \bibnamefont {Froud}}, \bibinfo
  {author} {\bibfnamefont {I.~C.~E.}\ \bibnamefont {Turcu}}, \bibinfo {author}
  {\bibfnamefont {P.}~\bibnamefont {Villoresi}}, \bibinfo {author}
  {\bibfnamefont {W.~A.}\ \bibnamefont {Bryan}}, \bibinfo {author}
  {\bibfnamefont {E.}~\bibnamefont {Springate}}, \ and\ \bibinfo {author}
  {\bibfnamefont {L.}~\bibnamefont {Poletto}},\ }\href {\doibase
  10.1364/OE.19.019169} {\bibfield  {journal} {\bibinfo  {journal} {Opt.
  Express}\ }\textbf {\bibinfo {volume} {19}},\ \bibinfo {pages} {19169}
  (\bibinfo {year} {2011})}\BibitemShut {NoStop}%
\bibitem [{\citenamefont {Grazioli}\ \emph {et~al.}(2014)\citenamefont
  {Grazioli}, \citenamefont {Callegari}, \citenamefont {Ciavardini},
  \citenamefont {Coreno}, \citenamefont {Frassetto}, \citenamefont {Gauthier},
  \citenamefont {Golob}, \citenamefont {Ivanov}, \citenamefont
  {Kivim{\"{a}}ki}, \citenamefont {Mahieu}, \citenamefont {Bu{\v{c}}ar},
  \citenamefont {Merhar}, \citenamefont {Miotti}, \citenamefont {Poletto},
  \citenamefont {Polo}, \citenamefont {Ressel}, \citenamefont {Spezzani},\ and\
  \citenamefont {{De Ninno}}}]{Grazioli:2014}%
  \BibitemOpen
  \bibfield  {author} {\bibinfo {author} {\bibfnamefont {C.}~\bibnamefont
  {Grazioli}}, \bibinfo {author} {\bibfnamefont {C.}~\bibnamefont {Callegari}},
  \bibinfo {author} {\bibfnamefont {A.}~\bibnamefont {Ciavardini}}, \bibinfo
  {author} {\bibfnamefont {M.}~\bibnamefont {Coreno}}, \bibinfo {author}
  {\bibfnamefont {F.}~\bibnamefont {Frassetto}}, \bibinfo {author}
  {\bibfnamefont {D.}~\bibnamefont {Gauthier}}, \bibinfo {author}
  {\bibfnamefont {D.}~\bibnamefont {Golob}}, \bibinfo {author} {\bibfnamefont
  {R.}~\bibnamefont {Ivanov}}, \bibinfo {author} {\bibfnamefont
  {A.}~\bibnamefont {Kivim{\"{a}}ki}}, \bibinfo {author} {\bibfnamefont
  {B.}~\bibnamefont {Mahieu}}, \bibinfo {author} {\bibfnamefont
  {B.}~\bibnamefont {Bu{\v{c}}ar}}, \bibinfo {author} {\bibfnamefont
  {M.}~\bibnamefont {Merhar}}, \bibinfo {author} {\bibfnamefont
  {P.}~\bibnamefont {Miotti}}, \bibinfo {author} {\bibfnamefont
  {L.}~\bibnamefont {Poletto}}, \bibinfo {author} {\bibfnamefont
  {E.}~\bibnamefont {Polo}}, \bibinfo {author} {\bibfnamefont {B.}~\bibnamefont
  {Ressel}}, \bibinfo {author} {\bibfnamefont {C.}~\bibnamefont {Spezzani}}, \
  and\ \bibinfo {author} {\bibfnamefont {G.}~\bibnamefont {{De Ninno}}},\
  }\href {\doibase 10.1063/1.4864298} {\bibfield  {journal} {\bibinfo
  {journal} {Review of Scientific Instruments}\ }\textbf {\bibinfo {volume}
  {85}},\ \bibinfo {pages} {23104} (\bibinfo {year} {2014})}\BibitemShut
  {NoStop}%
\bibitem [{\citenamefont {Escher}\ \emph {et~al.}(2011)\citenamefont {Escher},
  \citenamefont {Weber}, \citenamefont {Merkel}, \citenamefont {Plucinski},\
  and\ \citenamefont {Schneider}}]{Escher:2011}%
  \BibitemOpen
  \bibfield  {author} {\bibinfo {author} {\bibfnamefont {M.}~\bibnamefont
  {Escher}}, \bibinfo {author} {\bibfnamefont {N.~B.}\ \bibnamefont {Weber}},
  \bibinfo {author} {\bibfnamefont {M.}~\bibnamefont {Merkel}}, \bibinfo
  {author} {\bibfnamefont {L.}~\bibnamefont {Plucinski}}, \ and\ \bibinfo
  {author} {\bibfnamefont {C.~M.}\ \bibnamefont {Schneider}},\ }\href {\doibase
  10.1380/ejssnt.2011.340} {\bibfield  {journal} {\bibinfo  {journal}
  {e-Journal of Surface Science and Nanotechnology}\ }\textbf {\bibinfo
  {volume} {9}},\ \bibinfo {pages} {340} (\bibinfo {year} {2011})}\BibitemShut
  {NoStop}%
\bibitem [{\citenamefont {Ebert}\ \emph {et~al.}(2011)\citenamefont {Ebert},
  \citenamefont {K{\"{o}}dderitzsch},\ and\ \citenamefont
  {Min{\'{a}}r}}]{Ebert:2011}%
  \BibitemOpen
  \bibfield  {author} {\bibinfo {author} {\bibfnamefont {H.}~\bibnamefont
  {Ebert}}, \bibinfo {author} {\bibfnamefont {D.}~\bibnamefont
  {K{\"{o}}dderitzsch}}, \ and\ \bibinfo {author} {\bibfnamefont
  {J.}~\bibnamefont {Min{\'{a}}r}},\ }\href {\doibase
  10.1088/0034-4885/74/9/096501} {\bibfield  {journal} {\bibinfo  {journal}
  {Reports on Progress in Physics}\ }\textbf {\bibinfo {volume} {74}},\
  \bibinfo {pages} {96501} (\bibinfo {year} {2011})}\BibitemShut {NoStop}%
\bibitem [{\citenamefont {Ebert~et al.}(2017)}]{Ebert:2017}%
  \BibitemOpen
  \bibfield  {author} {\bibinfo {author} {\bibfnamefont {H.}~\bibnamefont
  {Ebert~et al.}},\ }\href {http://olymp.cup.uni-muenchen.de/ak/ebert/SPRKKR}
  {{\bibinfo {title} {{\textit{The Munich SPR-KKR Package}, version 7.7, http://olymp.cup.uni-muenchen.de/ak/ebert/SPRKKR}}}\ } (\bibinfo
  {year} {2017})\BibitemShut {NoStop}%
\bibitem [{\citenamefont {Braun}\ \emph {et~al.}(2018)\citenamefont {Braun},
  \citenamefont {Min{\'{a}}r},\ and\ \citenamefont {Ebert}}]{Braun:2018}%
  \BibitemOpen
  \bibfield  {author} {\bibinfo {author} {\bibfnamefont {J.}~\bibnamefont
  {Braun}}, \bibinfo {author} {\bibfnamefont {J.}~\bibnamefont {Min{\'{a}}r}},
  \ and\ \bibinfo {author} {\bibfnamefont {H.}~\bibnamefont {Ebert}},\ }\href
  {\doibase https://doi.org/10.1016/j.physrep.2018.02.007} {\bibfield
  {journal} {\bibinfo  {journal} {Physics Reports}\ }\textbf {\bibinfo {volume}
  {740}},\ \bibinfo {pages} {1} (\bibinfo {year} {2018})}\BibitemShut {NoStop}%
\bibitem [{\citenamefont {Johannsen}\ \emph {et~al.}(2015)\citenamefont
  {Johannsen}, \citenamefont {Ulstrup}, \citenamefont {Crepaldi}, \citenamefont
  {Cilento}, \citenamefont {Zacchigna}, \citenamefont {Miwa}, \citenamefont
  {Cacho}, \citenamefont {Chapman}, \citenamefont {Springate}, \citenamefont
  {Fromm}, \citenamefont {Raidel}, \citenamefont {Seyller}, \citenamefont
  {King}, \citenamefont {Parmigiani}, \citenamefont {Grioni},\ and\
  \citenamefont {Hofmann}}]{Johannsen:2015}%
  \BibitemOpen
  \bibfield  {author} {\bibinfo {author} {\bibfnamefont {J.~C.}\ \bibnamefont
  {Johannsen}}, \bibinfo {author} {\bibfnamefont {S.}~\bibnamefont {Ulstrup}},
  \bibinfo {author} {\bibfnamefont {A.}~\bibnamefont {Crepaldi}}, \bibinfo
  {author} {\bibfnamefont {F.}~\bibnamefont {Cilento}}, \bibinfo {author}
  {\bibfnamefont {M.}~\bibnamefont {Zacchigna}}, \bibinfo {author}
  {\bibfnamefont {J.~A.}\ \bibnamefont {Miwa}}, \bibinfo {author}
  {\bibfnamefont {C.}~\bibnamefont {Cacho}}, \bibinfo {author} {\bibfnamefont
  {R.~T.}\ \bibnamefont {Chapman}}, \bibinfo {author} {\bibfnamefont
  {E.}~\bibnamefont {Springate}}, \bibinfo {author} {\bibfnamefont
  {F.}~\bibnamefont {Fromm}}, \bibinfo {author} {\bibfnamefont
  {C.}~\bibnamefont {Raidel}}, \bibinfo {author} {\bibfnamefont
  {T.}~\bibnamefont {Seyller}}, \bibinfo {author} {\bibfnamefont {P.~D.~C.}\
  \bibnamefont {King}}, \bibinfo {author} {\bibfnamefont {F.}~\bibnamefont
  {Parmigiani}}, \bibinfo {author} {\bibfnamefont {M.}~\bibnamefont {Grioni}},
  \ and\ \bibinfo {author} {\bibfnamefont {P.}~\bibnamefont {Hofmann}},\ }\href
  {\doibase 10.1021/nl503614v} {\bibfield  {journal} {\bibinfo  {journal} {Nano
  Letters}\ }\textbf {\bibinfo {volume} {15}},\ \bibinfo {pages} {326}
  (\bibinfo {year} {2015})}\BibitemShut {NoStop}%
\end{thebibliography}
%
\end{document}